\documentclass[12pt]{amsart} 
\usepackage{amsmath,amsthm,amsfonts,amscd,amsxtra,amsopn,verbatim,array}
\usepackage{eucal}
\usepackage{epsfig}
\usepackage{amsfonts}
\usepackage{amsmath}
\usepackage{amsthm}
\usepackage{pdfpages}

\usepackage{mathrsfs}
\usepackage[all]{xy}

\renewcommand{\mathfrak}{\mathcal}
\renewcommand{\mathcal}{\bold}

\newtheorem*{thm}{Theorem}
\newtheorem*{prop}{Proposition}

\usepackage[numbers]{natbib}

\usepackage[colorlinks=false,linktocpage=true]{hyperref}

\newcommand{\beq}{\begin{equation}}
\newcommand{\eeq}{\end{equation}}
\newcommand{\beqarr}{\begin{eqnarray}}
\newcommand{\eeqarr}{\end{eqnarray}}
\newcommand{\beqa}{\begin{eqnarray*}}
\newcommand{\eeqa}{\end{eqnarray*}}

\begin{document}

\title[A discrete Kermack-McKendrick model]{Back to the roots: A discrete Kermack-McKendrick model adapted to Covid-19}
\author[\small M. Kreck, E. Scholz]{\small  Matthias Kreck$^{\dag}$,  Erhard Scholz $^{\ddag}$}
\date{\today}

\begin{abstract} 
{\tiny A widely used tool for analysing the Covid-19 pandemic is the standard SIR model. It seems often to be used as a black box, not taking into account that this model was derived as a special case of the seminal Kermack-McKendrick theory from 1927. This is our starting point.  We explain the setup of the Kermack-McKendrick theory (passing to a discrete approach) and use medical  information for specializing to  a model { called by us an {\em adapted K-McK-model}.  It  includes} effects of vaccination, mass testing and mutants. We demonstrate the use of the model by applying it to the development in Germany { and show, among others things, }
that a comparatively mild intervention reducing the time until quarantine by one day leads to a drastic improvement. 
}
 \end{abstract}

\maketitle
\renewcommand{\thefootnote}{\fnsymbol{footnote}}

\footnotetext[2]{Mathematisches Institut der Universit\"at Bonn and Mathematisches Institut der Universit\"at Frankfurt, Germany, \quad kreck@math.uni-bonn.de}
\footnotetext[3]{University of Wuppertal, Faculty of  Math./Natural Sciences, and Interdisciplinary Centre for History and Philosophy of Science, \quad  scholz@math.uni-wuppertal.de}
\renewcommand{\thefootnote}{\arabic{footnote}}

\setcounter{tocdepth}{1}

\section{Introduction}



{ Nearly a decade ago Breda/Diekmann et al.  made the sharp remark  
  that even experienced experts seem often to believe that Kermack and McKendrick's famous paper of 1927 \cite{Kermack/McKendrick:1927} is just about the  standard SIR or  SEIR model described by the well known ODE system  \cite[p. 105]{Breda/Diekmann:2012}. 
  }
Taking this criticism seriously, the present paper proposes a Kermack-McKendrick type model adapted to Covid-19,  which differs from the  widely used S(E)IR models and which { in our eyes} 
 has a greater plausibility and applicability { than the latter}.

 For doing so we work with a discrete time parameter. After all, people are not participating  in the spread of the disease during  every part of the day, e.g.  when they are asleep. And also the data are communicated on a daily basis. In addition, discrete models are easy to program. So we adapt the Kermack-McKendrick approach and replace functions depending on a continuous time $t$  by a functions depending on integers $k$. 

We will describe the model in the next section and then explain the role of its parameters (sections 1 and 2). Then we describe how one applies the model starting from the empirical data and demonstrate the result for Germany (sections 3 and 4).  We give a striking application showing what happens if people are sent earlier to quarantine or hospital (section 5), before we discuss how vaccination, the rise of new mutants, and mass testing can be implemented in the model
(section  6).  Finally, we add some remarks on the relationship between the reproduction numbers as they appear in our model to the ones given by the Robert-Koch-Institut   for Germany (section 9). At the end we discuss some aspects of our model.

\section{The model}
We recall the often neglected fundamental 
fundamental idea 
 of Kermack and McKendrick following \cite{Kermack/McKendrick:1927,Brauer:2017,Breda/Diekmann:2012}. 
As {already} mentioned 
we use the discrete setting of the  Kermack-McKendrick approach, which the original paper started with before the  continuous version  was introduced as a limiting case. A central input of a Kermack/McKendrick model is a function $\gamma(j)$ that measures the medical strength of infectivity of a person at day $j$ counted from 
{ 
the onset  of infection.}
 On this basis the Kermack-McKendrick approach allows one to derive equations which express how many people are susceptible to the virus if one knows the mean contact rate of the interaction of the population. To emphasize that $\gamma$  is the central input we call such a model a {\em discrete $\gamma$-K-McK model}. 
 {
 A general discussion of discrete K-McK models can be found in the recent paper of Diekmann et al.  
 \citep{Diekmann_ea:2021}.
 }

If you want to construct a $\gamma$-K-McK-model the first thing to do is to look for virological studies which determine $\gamma (j)$ as closely as possible. This will depend on the type of an epidemic. The closer the input function $\gamma(j)$ is to the medical facts, the more reliable is the model.
{  Combining the information of this function with estimates of time-dependent contact and transmission rates, which we propose to determine from the data of daily  registered newly infected people (see section \ref{section data to model}), and  the parameters $e, p_c$ explained below, we are able  
} to model the number of susceptibles 
{ and the number of}
newly infected at day $k$. 

After infection there are a few days where people are exposed, meaning that they are already infected but not yet infectious. Then the period begins 
{ in which}
a person is infectious (we call 
{ these people}
propagators) and after that {  may, in general,} be considered as immune. The period of both lengths is not sharp { and varies a bit from person to person}. Our assumption in this paper is that this variation is small, meaning that the functions that  describe these lengths are nearly  jump functions. In other words we assume that  after infection there are $e$ days, where people are exposed 
{ 
to the virus}
but not yet infectious. After that there are $p_d$ days where a person is infectious and after that in general may be considered as immune. There are indications that this is the case for Covid-19. %
Amongst those who are infectious there are those who are positively tested after $p_c < p_d$ days on average, and then are confined to home isolation or hospital. (We will explain the notation $p_d$ and $p_c$ later, for an example see fig. 1.) Functions $\gamma$ with only a finite number of days where they are non-zero are called {\em functions with finite support}.
 
The $\gamma$-K-McK-models are flexible enough to allow  for adding important additional features  of an epidemic such as, e.g., unreported cases, vaccination, effects of testing and of the rise of new mutants etc. Here we take up the first point only and discuss the other points later.  

The assumption of a finite support of $\gamma$ fits with the following picture: At each day the population  can, in the model, be divided into disjoint subsets, 
called compartments, as follows:
\begin{itemize}
\item[--] The {\em compartment $S$  of susceptible people} to the disease with cardinality $S(k)$ at the day $k$. Here susceptible means not immunised, i.e. neither recovered nor vaccinated. 
Here k is in $\mathbb Z$ and we assume that $S(k) = N$, the total number of the population, for $k$ smaller then some negative bound. The time between this negative bound and day $0$ is the prehistory. We will give a renewal equation for $S(k)$ which uses as an input a prehistory. 
\item[--] The {\em compartment $E$ of exposed people} whose cardinality at the day $k$ we denote by $E(k)$. By this we mean people who are infected but are not yet infectious. The function $\gamma$ determines the number $e$ of days people stay in compartment $E$ as 
\[
e:= \text {min} \, \{j \,|\gamma (j) \ne 0   \} - 1 \; .
\]
\item[--] The {\em compartment $P$ of propagators}. These are people who are infectious (they propagate the virus) but not yet in quarantine or isolation. We will divide this into the following subcompartments: 

\noindent
The {\it compartment $P_c$ of ``counted" propagators}  with cardinality $P_c(k)$, these are  the infectious who will later be diagnosed, recorded and counted in the statistics. As mentioned above the number of days people stay in this compartment is denoted by $p_c$. 

\noindent
And the {\it compartement $P_d$ of} 
{\em propagators}  who are never reported as infected, {\em the dark sector} with cardinality $P_d(k)$.  The number of  days people stay in this compartment is again derived from $\gamma $ as the cardinality of its support:
\[
p_d:= \sharp \, \{j \,|\gamma (j) \ne 0   \} 
\]
\item[--] The {\em compartment $Q$} of  persons who are diagnosed and {\em quarantined} or isolated  until recovery or death.  The cardinality of this compartment is  $Q(k)$ and  the   mean residence time  is denoted by $q$. The members of $Q$  are   essentially {\em no longer infecting others}, although they are still infectious in the medical sense. The number $Q_{\text{new}}(k)$ of daily new registered persons in $Q$ is a central datum for the epidemic.
\item[--] The {\em compartment $R$ of removed} from the epidemic (recovered or dead) with cardinality $R(k)$. 
\end{itemize}

For modelling an epidemic realistically in the $\gamma$-K-McK framework which also includes the dark sector  and accounts for changing contact rates  one needs other input parameters which have to be estimated from empirical data: 
\begin{itemize}
\item[--] We assume that  for the infected people leaving compartment $E$ at any day $k$ there is a  a certain fraction $\alpha(k)$ of people from $E$ who move to compartment $P_c$  at
 day $k$, whereas the fraction $ 1 - \alpha(k)$ of people moves to the compartment $P_d$ of people in the dark sector. 
\item[--] It could be that people in the dark sector are less infectious by a factor $\xi \le 1$, so this has to be taken into consideration. A lot of them have no symptoms  which might indicate that they have lower viral load. 
\item[--]
 Besides the medical function $\gamma(j)$ for a certain epidemic one needs to know a daily proportionality factor  called for simplicity the {\em contact rate} $\kappa(k)$. 
\end{itemize}

Now we explain the {\em dynamics of the model}. 
The idea is that all people in compartments $P_c$ and $P_d$ infect people from compartment $S$ proportionally to the strength of the infection given by $\gamma$ and the contact rate $\kappa(k)$. Here we note that both functions $\gamma$ and $\kappa$ are dependent on the choice of a unit, but if we replace $\gamma$ by $c \, \gamma$ we have to compensate for this by replacing $\kappa$ by $\frac \kappa c$. Another convention influencing $\kappa$ is to introduce $s(k) : = \frac{S(k)}{N}$, where $N$ denotes the size of the population, which in this paper we consider as constant.

To explain the dynamics in detail we denote the number of people who entered compartment $E$ for the first time on day $k$ by $E_{\text{new}}(k)$. In the situation when no vaccination is 
available (we shall discuss the variants with vaccination in Section \ref{section vaccinations}) it  can be expressed in terms of the function $S$ as:
\beq
E_{\text{new}}(k) = S(k-1) -S(k) \,. \label{eq Enew1}
\eeq
Similarly, we denote the number of people who  entered compartment $P_c$ on day $k$ for the first time by $P_{c,new}(k)$ and similarly those who   entered compartment $P_d$ on day $k$ for the first time by $P_{d,new}(k)$. We first explain the dynamics of those who  are  in compartment $P_c$ at the day $k-1$. They are the disjoint union of people who newly entered this compartment during the $p_c$ days before. So for each $1 \le j \le p_c$ there are $P_{c,new}(k-j)$ people in $P_c$, who infect people who are in $S$ at  day $k-1$  with strength $\gamma (j)$ and proportional to the contact rate $\kappa (k-1)$. 

From $\kappa$ and $\gamma$ one can read off the probability that a person who entered compartment $P_c$ at day $k-j$ infects a susceptible person at day $k-1$, namely as $\frac {\kappa (k-1) \gamma (j)} N$.  This implies that the overall probability involving all people who entered compartment $P_c$ at day $k-j$ to infect a susceptible person at day $k-1$ is \cite[p.~4]{Diekmann_ea:2021}
\beq 1 - \Big(1-\frac {\kappa (k-1) \gamma (j)} N\Big)^{P_{c,new}(k-j)}\, . \label{eq prob infectivity Odo}
\eeq
Often the linear approximation 
\beq \frac {\kappa (k-1) \gamma (j)} N \, {P_{c,new}(k-j)}  \label{eq linearisation infectivity}
\eeq 
is used instead. In our situation the difference is negligible and so we use the linearisation (\ref{eq linearisation infectivity}). The contribution to the new infections on day $k-1$ by those people who entered compartment $P_c$ at day $k-j$ is  then (for $1 \leq j \leq p_c$).
$$s(k-1)\kappa (k-1) \gamma (j)  P_{c,new} (k-j).
$$
Similarly for the compartment $P_d$ (for $1 \leq j \leq p_d$) the contribution is 
$$\xi s(k-1)\kappa (k-1) \gamma (j) P_{d,new} (k-j).
$$
 Summing  up we obtain the following equation expressing the dynamics of the epidemic: 
\begin{equation}
E_{\text{new}} (k) =  s(k-1)\, \kappa (k-1) \, \biggl [\, \sum_{j=1} ^{p_c} \gamma (j+e) P_{c,new} (k-j) + \xi \sum_{j=1} ^{p_d}  \gamma (j+e) P_{d,new} (k-j)\, \biggr] \label{eq Enew2}
\end{equation} 
If we were working in a continuous model, the sum would be an
integral from $0$ to ~$p_c$.

{ To derive 
a single recursion equation for the number of susceptible people at day $k$ from these inputs  we  express} $P_{c,new} $ and $P_{d,new}$ in terms of  $S$. 
{
As the newly exposed people at day $k$ move to compartments $P_c$ and $P_d$ after $e$ days  with ratio $\alpha$ respectively $1- \alpha$ one finds:}
\beq
P_{c,new}(k+e) = \alpha (k+e)  E_{\text{new}}(k) = \alpha (k+e)  \bigl (S(k-1) -S(k)\bigr)\,  \label{eq Pc new}
\eeq
\beq
P_{d, new}(k+e)= (1-\alpha (k+e)  ) E_{\text{new}}(k) = \bigl (1-\alpha (k+e) \bigr  )\bigl(S(k-1) -S(k)\bigr)\,  \label{eq Pd new}
\eeq

To simplify the presentation we introduce two functions, which summarise the two expressions on the right hand side of equation (\ref{eq Enew2}{)}. 
\begin{equation}
\mathbb P _c (k-1) := \sum_{j=1} ^{p_c} \gamma (e+j)\,  \alpha (k-j)\,  \bigl[S(k-j-e-1) -S(k-j-e)\bigr]\, , \label{eq Pc}
\end{equation}
 \begin{equation}
\mathbb P_d (k-1) :=  \sum_{j=1} ^{p_d}  \gamma (e+j)\, \bigl(1- \alpha (k-j)\bigr) \, \bigl[S(k-j-e-1) -S(k-j-e)\bigr] \, .\label{eq Pd}
\end{equation}
{
  $\mathbb P _c $ and $\mathbb P _d$ express the contributions of the  propagating infected people who have been counted, $P_c$, and of the uncounted propagating individuals of the dark sector, $P_d$, respectively, and are not to  be confused with the latter.
}
Using this and equations (\ref{eq Enew1}) - (\ref{eq Pd}) we have finished the derivation of our model equation. \\

\newpage
\noindent
\begin{prop}  Let the functions $\gamma$, $\kappa$, $\alpha$ and the integer $p_c$ be given. 
We further set $S(k) = N$ for $k < -(p_d + e)$ and chose values $S(k)$ (a prehistory) for values $-(p_d +e) \le k \le 0$. Then for $k>0$, if the dynamics of an epidemic is given by equ.  
 (\ref{eq Enew2}), the single recursive equation for the number of susceptible people is
\begin{equation}
S(k-1) - S(k)  = s(k-1)\, \kappa (k-1)\,  \bigl[\mathbb P_c (k-1)
+ \xi\, \mathbb P_d (k-1)\bigr] \,. \label{eq recursion}
\end{equation}\\
The functions  $E$, $P_c$, $P_d$, and $Q$ are expressed in terms of $S$ as follows:
\begin{equation}
E(k) = S(k-e) - S(k)
\end{equation} 
\begin{equation}
P_c(k) = \sum_{j=0}^{p_c-1}\alpha(k-j)\, \bigl(S(k-e-j-1) - S(k-e-j) \bigr) \label{eq P_c}
\end{equation} 
Similarly 
\begin{equation}
P_d(k) =  \sum_{j=0}^{p_d-1}\bigl(1-\alpha(k-j)\bigr) \, \bigl(S(k-e- j-1) - S(k-e-j) \bigr) \label{eq P_d}
\end{equation} 
and
\begin{equation}
Q(k) = \sum_{j=0}^{q-1}\alpha(k-p_c- j) \,\bigl(S(k-e-p_c- j-1) - S(k-e-p_c-j) \bigr). \label{eq Q}
\end{equation} 
The number of reported as newly infected people, denoted by $Q_{\text{new}}(k)$ is
 \begin{equation}
Q_{\text{new}}(k) = \alpha(k-p_c) \,  \bigl(S(k-e-p_c-1) - S(k-e-p_c)\bigr)\, . \label{eq Q_new}
\end{equation}
\end{prop}

{
\vspace{1em}
\begin{proof}
We have given the proof of the recursion equation \ref{eq recursion} before the proposition. 

 Since people stay in compartment $E$  for $e$ days before they move on to the compartment $P$ we obtain:
$$
E(k) = E_{\text{new}}(k) + E_{\text{new}}(k-1) + \dots + E_{\text{new}}(k-e+1) 
$$
$$ = - \bigl(S(k) - S(k-1)\bigr) + \bigl(S(k-1) - S(k-2)\bigr) + \dots +\bigl (S(k-e+1) - S(k-e)\bigr)$$
$$
= S(k-e) - S(k)
$$
If we apply   the same summation to the recognized propagating people, 
\[
P_c(k) = P_{c,new}(k) + P_{c,new}(k-1) + \dots +P_{c,new}(k-p_c+1) \, ,
\]
then one obtains (\ref{eq P_c}), and the proof of (\ref{eq P_d}) is the same. The proof of (\ref{eq Q}) is also the same, once we have formula (\ref{eq Q_new}). For this we note that people  entering compartment $Q$ on day $k$ for the first time are the new people entering the  compartment $P_c$ at day $k-p_c$ the  compartment $P_c$: 
$$
Q_{\text{new}}(k) = P_{c,new}(k-p_c)
$$
and so by formula (\ref{eq Pc new}) we  have
$$
Q_{\text{new}}(k) =  \alpha (k-p_c)  \bigl (S(k-e-p_c-1) -S(k-e-p_c)\bigr)
$$ 
\end{proof}

\noindent 
We will give a more general version of our model involving the effects of vaccination, testing a certain percentage of the population each day and of arising of new mutants in section \ref{section vaccinations} (formula \ref{eq recursion test}).

An important key figure of an epidemic is the daily reproduction number $\rho(k)$. This is  defined as the average number of secondary infected people by one typical primary infected person  (averaged over all infectious at the day k). Since there are no data available from which one can directly read off the reproduction number the best thing one can do is to derive it from a reliable model. \\

\noindent
 {\em 
The  reproduction numbers  in the adapted K-McK-model are:
\begin{eqnarray}
\rho(k) &=& \sum_{j=0}^{p_c-1} s(k-1+j)\kappa(k-1+j)\alpha(k-1+j)\, \gamma(j+1)   \label{eq rho}\\
 & & \quad + \;\sum_{j=0}^{p_d-1} s(k-1+j)\kappa(k-1+j)(1- \alpha(k-1+j))\, \gamma(j+1)\nonumber  
\end{eqnarray}
}
\noindent For approximately constant values of $\alpha(k), s(k), \kappa(k)$ in the interval $[k, k+p_d)$ this boils down to a weighted average of the number of secondary infected by  a person entering $P_c$ at day $k$ and the analogous number for those entering $P_d$ at the same day. 

\noindent
{\bf Remark:} {\em The model is characterized by a delay structure which  appears  typically in  Covid-19. For example, a change in the contact rates $\kappa$ at a day $k$ will                                                                                                                                                                                                                                                                                                                                                                                                                           be observable in the  numbers of daily new recorded infected $Q_{\text{new}}$ only $e+p_c$ later (a sharp  eye will notice the  time delay in fig.~\ref{fig Anew D}).
}  

Similar models without a dark sector and with a simple box-like function $\gamma$ were constructed and used in \cite{Mohring_ea:2020,Fessler:2020,Shayak/Sharma:2020,K-S-2}. 

There are other discrete time models which are similar to the present one in spirit, but differ in detail. For example, after our paper was submitted we became aware of an interesting paper by Sofonea et al.,  which also develops a discrete-time model based on the force of infection \citep{Sofonea_ea:2021}. 
A central difference of this paper is that the authors are interested in the development in France where data about new infections are not reliable. So they aim for an analysis of ICU activity and hospital mortality time-series, whereas we aim for an analysis of the number of newly infected as link to real data. To do this the authors introduce new compartments which on the one hand distinguish different age groups, and on the other hand distinguish between non-critical infectious, critical infectious, long-stay ICU hospitalised, other critical hospitalised patients, recovered immunised, dead (cumulative mortality) and (like we) latent. This complicates the formulas but the basic idea of the dynamics is very similar. Our equation (\ref{eq recursion}) corresponds to their first equation in the supplement S2.1: $S_{i+1} = S_i - \Lambda_i S_i$, where $\Lambda_i$ is -- in a slightly different interpretation -- the force of infection. Besides involving the different compartments of infectious people the difference is that our function $\gamma$ is replaced by the generation time distribution $\zeta$ (sometimes the authors call it the serial time distribution). This is a different perspective. While we stress the medical strength of the virus, the authors relate the force of infection to the time between the infection of an infector and the infection of his or her infectee, which expresses the epidemiological perspective. But the role it plays for the model dynamics is the same. We will return to this point later when we discuss the determination of $\gamma$. Another difference is the way the authors count contact rates. For example if one reduces their model to the case of a single age group and a single group of infectious people our contact rate $\kappa$  corresponds to the square-root of their contact rate $c$. There are a few other small differences which look like ad hoc formulations, for example the use of a Michaelis-Menten formula in equation (S-4) rather than (\ref{eq prob infectivity Odo}). The linearisation  used by the authors in practice is the same as our (\ref{eq linearisation infectivity}).

 \section{ The role of the parameters \label{section parameters}}
If one wants to model an epidemic one has to find the appropriate parameters. If, for the sake of argument, we remove the dark sector for a moment,   there are  two types of parameters in the model:  parameters that are of a medical  nature such as the duration of the exposed people period $e$ and the infectivity function $\gamma(k)$,  and those that can be influenced by politicians,  such as the duration $p_c$ between the onset of infectivity and the beginning of quarantine or isolation, or the daily contact rate $\kappa(k)$. Coming back to the dark sector, one has to know in addition the ratio $\alpha$ between the  later recorded  people and the rest, and the duration $p_d$ of the infectious phase of the dark people. Amongst these parameters we assume that $e$ and the strength function $\gamma$ are essentially unchanged during the course of the epidemic, as well as $p_c$, $p_d$ and even $\alpha$ if we abstract from changes in the  ratio due to rising unspecific testing. The parameter which definitely changes over time is $\kappa(k)$.

\noindent
{\bf Observation:} {\em If we start with data for $S(k)$ and keep all parameters fixed except $\kappa (k)$, then  $\kappa (k)$ is determined by equation (\ref{eq recursion}) as the quotient of the left hand side by the factor of $\kappa (k)$ on the right side. With these $\kappa (k)$ 
the data are identically reproduced by the model. 
 In turn if we start with arbitrary values of $\kappa(k)$ this gives us values $S(k)$ if we assume constant values for  $\alpha$. 
 So if we allow ourselves to read $\kappa (k)$  from equation (7) we obtain a bijection between values $S(k)$ and values $\kappa (k)$. 
  With  other words, if one allows daily changing values of $\kappa (k)$ derived from the empirical data the model is a tautology.}

This holds for any input function $\gamma$ whatsoever, also for continuous time $t$,  in particular for 
\begin{equation}
\gamma_{SIR} (t) = \lambda e^{-\mu t}  \label{eq SIR gamma}
\end{equation} 
\noindent 
 which is the input for the SIR model \cite{Breda/Diekmann:2012}. This changes if one assumes constant contact rates, or contact rates which are constant for some period and then are changed by a certain factor to a new constancy level. Then the models cease to be tautological.  In the section \ref{section comparison SIR} we will demonstrate this difference by comparing our model with the SIR model. 

Thus the choice of $\gamma (k)$ is essential for obtaining a realistic model for Covid-19. This is a delicate point since data from which one can derive $\gamma$ are not easy to access. From our point of view $\gamma$ reflects medical data, more precisely the viral load  and the probability of successful cell culture isolation (in the following ``culture probability''). From the point of view of the paper by Sofonea et al.  \citep{Sofonea_ea:2021}, which we shortly discussed above,  $\gamma$ is replaced  in their model by the serial time distribution $\zeta$. So it is useful to compare them. In addition we  will discuss a third way to obtain information about $\gamma$, which is just based on observations when the force of infection starts to leave the level $0$, when it obtains its maximum and when it decreases to $0$ again. 

We found two studies   \cite{Woelfel_ea:2020}, \cite{Jones/Drosten_ea:2021} by virologists which give very similar values for $\gamma$. Simplifying a bit, this leads  to the following picture. The length of $e$ is approximately $2$ days, so $\gamma (j) = 0$ for $1\le j \le 2$. Within another 2 days it  reaches a maximum close to the moment where symptoms show up, before it starts to decrease, slowly at the beginning and then faster until, after about further $10$ days,  it reaches a value where people are practically no longer infectious. This indicates that $p_d$ is $11$. Curves which show this increase and  decrease for 
  a small number of different infected people can be found in \cite{Woelfel_ea:2020}.  They allow one to 
 infer (to ``guess'') a typical course of the strength of infectivity $\gamma (k)$ for Covid-19 from the data on culture probabilities depending on the time after infection \cite[fig. 1.f]{Woelfel_ea:2020}.  A recent evaluation of the data of $25,381$ positive subjects comes to a similar course of positive culture probability \cite[fig. 4.c]{Jones/Drosten_ea:2021}.
  Using the information of these studies we 
  establish 
   the following table:
 
 \vspace{1em}
 \begin{center}
\begin{tabular}{| c | c | c | c | c | c | c | c | c|  c|  c | c|  c | c |  c| }
 \hline 
$k$ & 0 & 1 & 2 &3 &4 &5 &6 &7 &8 &9 &10&11&12 \\ 
\hline
$\gamma (e+k)$  & 0 &  0.5  & 0.9 & 0.9 &0.85 & 0.8 & 0.7& 0.6& 0.45& 0.15 & 0.05 & 0.02 & 0 \\
\hline
\end{tabular}
\end{center}
 \vspace{1em}
 \noindent
If one compares  this table for $\gamma$ with the values  for $\zeta$  displayed in figure 1 in \citep{Nishiura_ea:2020}  to which  Sofonea et al.  \citep{Sofonea_ea:2021} refer as reference for empirical data,  one observes a similar course.  

\newpage
We use a third way to obtain information about $\gamma$ which just uses the time span until the maximum is reached and the time needed to fall down  again to  zero after a short phase of staying close to the maximum. 
If one requires that such a transition is modelled by a function which is  arbitrary often differentiable, and at the interval ends all derivatives are zero,  one can  use  a natural standard candidate frequently used in mathematics as transition function. Normalized to the interval $[0,1]$  it is:
\beq
g(x) := \frac{ f(x)}{f(x)+  f(1-x)}\, , \label{eq g(t)}
\eeq
where
\[ f(x) = 
 \Big\{
{ e^{-\frac{1}{x}} \quad \mbox{for}\quad x>0   
\atop \; \; 0 \qquad \;   \mbox{for} \quad x \leq 0\, .}\;
\]

Using this function and the corresponding function for the decrease and adapting both to the data curve for Covid-19 one obtains another candidate for $\gamma$  expressed by
\beq \tilde{\gamma}(t)= 0.9 \, g(\frac{t-2}{2}) \big(1-g(\frac{t-4}{12})\big) \label{eq gamma-tilde}
\eeq 
 which is rather close to that obtained by the virological studies mentioned above, as figure \ref{fig gamma} (top right) shows.
Summarizing we come to the conclusion that, although it is impossible to derive $\gamma$ (or $\zeta$)  precisely, there are strong reasons to believe that our choice for $\gamma$ is not far off.
 We will later compare our model with the standard SIR model. Here we only compare our input function  $\gamma $  fig.~\ref{fig gamma} (top)  with the input function for the latter (bottom). 

\begin{figure}[h]
\begin{center}
\includegraphics[scale=0.5]{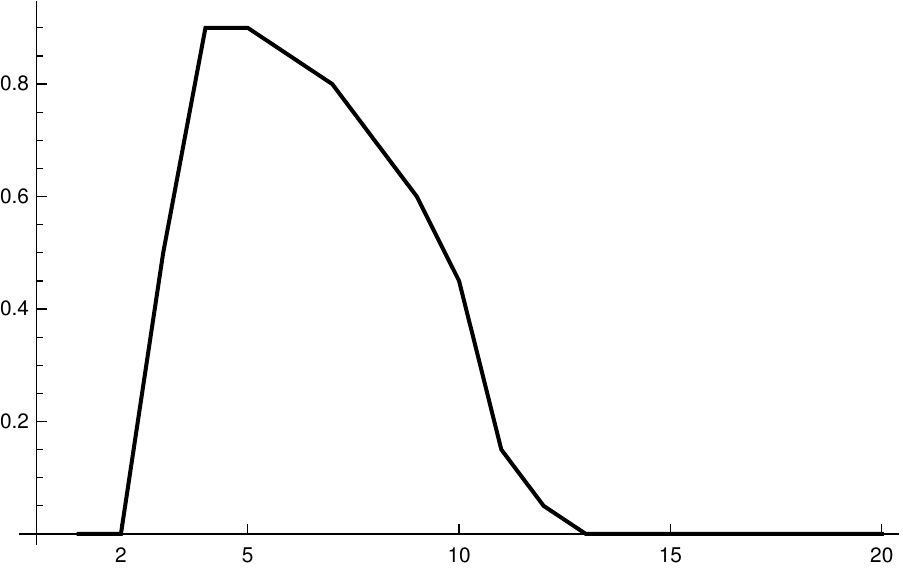} \qquad \includegraphics[scale=0.5]{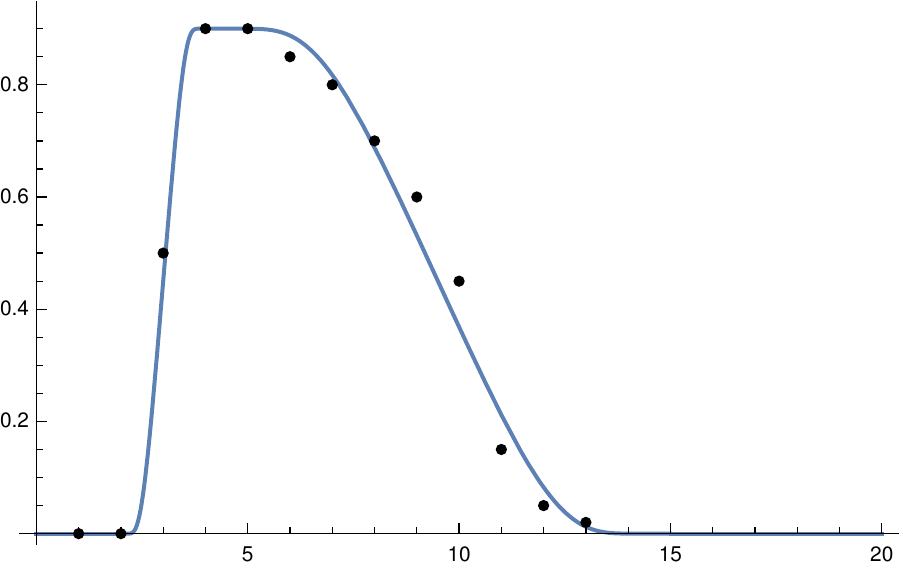} \\  \includegraphics[scale=0.5]{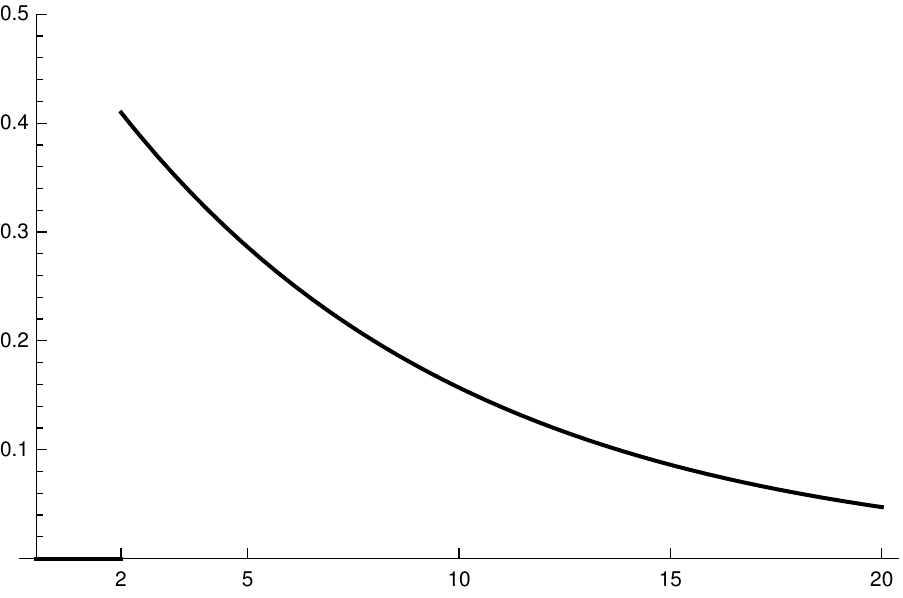} \qquad \includegraphics[scale=0.5]{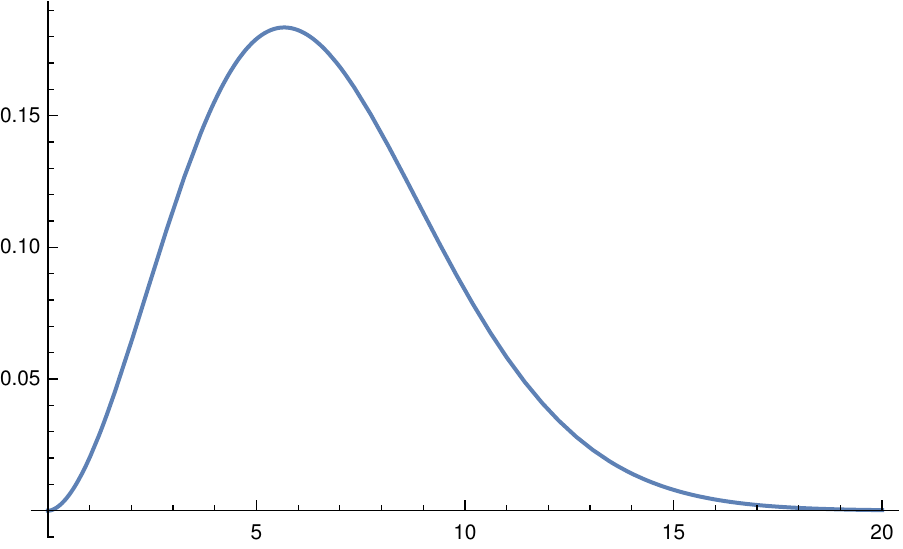}
\caption{\small Top left: $\gamma(j)$ of the adapted K-McK model, linearly interpolated. \quad Top right:  $\tilde{\gamma}(t)$ acccording to  (\ref{eq gamma-tilde}) (blue) with data points (black) from $\gamma$-table above.
Bottom left: Graph of $\gamma_{SIR}$ of the standard SIR model with values for $\lambda=0.41, \mu=0.12$ at the beginning of Covid-19 in Germany used in  \cite[fig.~1]{Dehning-ea:2020}. \quad Bottom right: Weibull distribution with parameters given in the main text.
In all four cases cut-off after 20 days.  \label{fig gamma}  }
\end{center}
\end{figure}

We remark that this is the point where one can see why in \cite{Breda/Diekmann:2012} the authors speak about an``incessant misconception'' if people apply the SIR-model indiscriminately  (and the situation is not so different for the SEIR model). The function  $\gamma (t) = \lambda e^{-\mu t}$  -- at least for Covid 19 -- has nothing to do with the medical data observed by the virologists.
This may have considerable consequences for conditional predictions; see section \ref{section comparison SIR}. 
{  Other possibilities for adapting the model to Covid-19 specific  data can be found in the literature. For example,  Ferretti et al. try to reconstruct the distribution of ``age'' (time after infection) dependent generation time from the analysis of data  of  40 source-recipient transmission pairs documented in the literature by May 2020 \cite{Ferretti_ea:2020}. They conclude that among several distributions used for  comparison the Weibull distribution with shape parameter $k=2.826$ and scale $\lambda=5.665$ fits best. 
The choice of the parameters suppresses the tail of the distribution for $t > 15$ strongly enough to distinguish it qualitatively from standard SIR distributions.
 It is difficult to make sure that the contact data used for the evaluation stem from comparable conditions of the social contact regime. It is not clear to us whether this condition is satisfied. In our approach, on the other hand, we rely on the medical data for the $\gamma$-function and include the effects of  social (and seasonal) transmission conditions by a proportionality factor $\kappa(t)$ determined from the data of recorded newly  infected people (see next section).
}

A model using the biological input function $\gamma$ above will be called a   K-McK-model adapted to Covid-19, or for short an {\em adapted K-McK-model}. To specify such a model one has  to estimate three further parameters $p_c$, $\alpha(k)$ and $\xi$. 

The next parameter one has to chose is $p_c$. 
It is difficult to estimate and depends on the country and the action mode of the health institutions. For Germany it is estimated to lie between 5 and 8; where for the year 2020 officials of the health system estimate it in the upper range. So we work with   $p_c = 7$. There are data that indicate that  the ratio $\alpha$ of {\em recorded} new infections among all new infections  was about $0.25$ for Germany at the beginning of the epidemic, but changed  to approximately $0.5$ during the course of the year, probably due to the increasing test rates \cite{Helmholtz-Zentrum-Muenchen:2020-12}. Since the influence of the dark sector was negligible in the first months we have chosen $\alpha =0.5$ for the whole period. Finally, it is difficult to estimate the relative strength   of infectivity   $\xi$ of the unrecorded infected people from empirical data, so we take $\xi =1$, i.e., we assume that people in the dark sector (including the asymptomatic ones) are basically as infectious as those in $P_c$. 

\section{ From data to model curves \label{section data to model}}

In this paper 
 we start from the data given  by the {\em Humdata} project of the Johns Hopkins University.\footnote{\url{https://data.humdata.org/dataset/novel-coronavirus-2019-ncov-cases}}  
 More precisely, we  estimate the values for $S(k)$ from the listed confirmed cases and the ratio $\alpha$. In this step  the estimated time translation $e+p_c$  between the times of infection and recording has to be taken into account. Moreover, there are weekly fluctuations of the values of newly recorded cases (the first differences of the data lists of confirmed cases) due to weekends where data taking  and transmission is usually slowed down considerably. Since we are interested in the central development of the epidemic we replace the daily values of the newly recorded by (centered) 7-days averages and also the values of the susceptible people $S_7(k)$.

In the next step we compute the daily contact rates $\kappa (k)$ according to the observation above. This allows one to give the tautological model picture of the values of $S_7(k)$ according to the observation above; but of course this is not what we want. We therefore investigate the values of $\kappa (k)$ and look for periods where they are more or less stationary and can be approximated by a constant. 
Such time intervals may be interpreted as  periods where the behaviour of the population is roughly the same and no crucial change of the virus occurs. The transition times between such stationary phases  seem to characterize periods in which the contact behaviour of the population changes. This is often the result of political measures (non-pharmaceutical interventions) reducing the number of contacts or allowing more contacts, which in many cases may lead to new  levels where the contact rates can again  be approximated by a constant. Besides this, eyeballing will indicate further levels of near-constancy that then need to be justified. For example, the explanation might be that the changed contact rate is the result of a climate change from summer to winter where people meet more in closed rooms, or the other way round. It could also be that  the infectivity of the virus changes (e.g.~with the rise of a more aggressive variant). In our view the most important criterion for the applicability of a Kermack/McKendrick type model to  a real-life epidemic boils down to the question of whether or  not replacing the contact rates during periods where they stay  nearly constant by their averages, or a value close to it inside the respective $\sigma$-interval, leads to a good approximation of the data curves. 

In the case of Germany, one observes seven rather obvious periods between March 2020 and mid-January 2021 in which $\kappa$ can be approximated by a constant, ignoring fluctuations (see fig.~\ref{fig kappa D}). Some of these constancy intervals can be interpreted as results of interventions, others not:  There was a series of three interventions in March 2020 resulting in the first constancy interval from ($t_1=$) March 24, 2020,   to ($t_2=$) April 26, a rather short interval. Around April 26 the interventions were reduced and the result is a long period until ($t_3=$) July 3. Already in this phase
 people were presumably getting  more careless, the vacation period started and  contacts increased, which lead to the third period, lasting until ($t_4=$)  September 27. In this phase the mean reproduction rate rose  noticeably above 1.  With the beginning of  cooler temperatures and  life moving more to closed rooms, the reproduction number rose to about 1.5 in the fourth interval until ($t_
 5=$) October 31. This provoked new containment measures, at first at Oct. 16 and Nov. 2  which are reflected by the fifth interval lasting until Nov. 26, in which the daily new infections went down. They started to rise again, probably because of early Christmas shopping on ($t_6=$) Nov. 26.
  A partial lockdown (schools, restaurants, cultural activities)  on December 16  (our $t_7$)  brought the reproduction rate  below 1 and resulted in, on average, falling numbers of new infections. This period lasts until the end of the data considered here in 
  mid-January (Jan. 15).

The  model values of the contact rates $\kappa_j$ in the intervals $[t_j, \, t_{j+1} -\Delta_{j+1}]$  ($1 \leq j \leq 7$), with varying durations $\Delta_j$ of the transition periods, are chosen close to the mean values, with small deviations inside the $\sigma$-interval allowed if this improves the fit considerably.
The model $\kappa_j$ and the (effective) model reproduction numbers $\rho_j$ at the left edge of the intervals are given in the following table:
\vspace{1cm}
\begin{center}
\begin{small}
\begin{tabular}{|l||c|c|c|c|c|c|c|}
\hline 
\multicolumn{8}{|c|}{\small $\kappa_0$ and model $\kappa_j$, $\rho_j$, $t_j$  for intervals $J_j$   for Germany}\\ 
\hline
 & $J_1$ & $J_2$ & $J_3$ & $J_4$ & $J_5$& $J_6$ & $J_7$  \\
\hline
 $\kappa_j$  & 0.131  & 0.162 & 0.208&  0.271  & 0.180  & 0.207 &0.164 \\
 $\rho_j$  & 0.73   & 0.90 & 1.16  & 1.50 & 0.99 & 1.12& 0.88  \\
$t_j$ (M/D) 2020 &  03/24 & 04/26 & 07/03 & 09/27 & 10/31 &  11/26 & 12/16 \\
\hline
\end{tabular}
\end{small}
\end{center}

\begin{figure}[h]
\begin{center}
\includegraphics[scale=0.615]{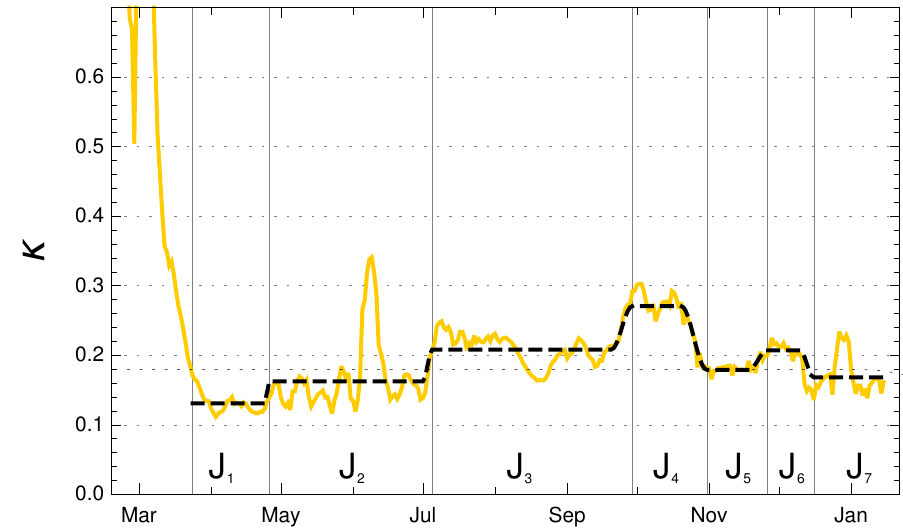} \hspace*{0.5em}\includegraphics[scale=0.6]{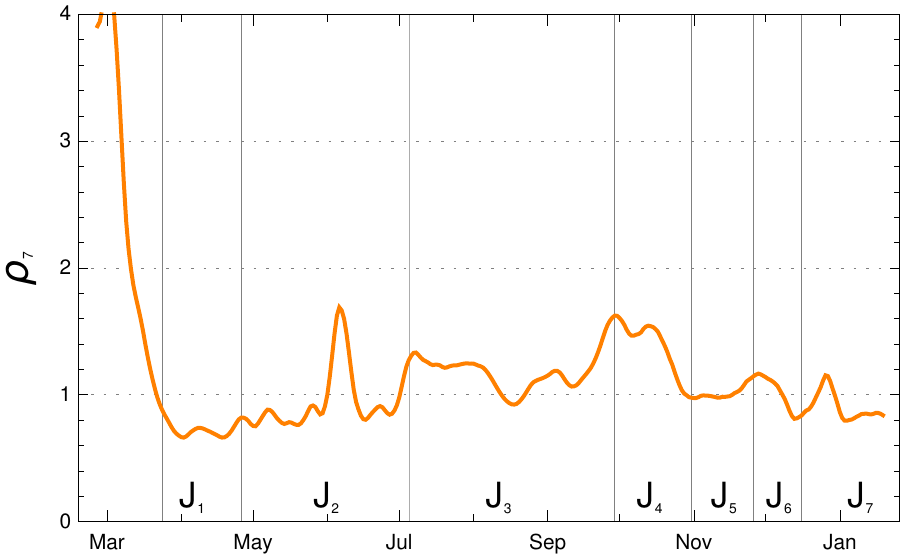} 
\caption{\small Left: Relative contact rates for Germany $\kappa(k)$ until mid-January 2021 (yellow) and model values in constancy intervals (black dashed), critical value corresponding to reproduction rate = 1 (dotted straight line). Right: Corresponding  reproduction rates. \label{fig kappa D}  }
\end{center}
\end{figure}

\vspace{1em}

Using this input for our recursion formula determines the model numerically. The values for the recorded daily new infections $Q_{\text{new}}$ are shown in fig. \ref{fig Anew D},  the numbers of recorded recovered people $R_c$ and of the recorded actual infected people $Q$  in figur  \ref{fig Rcrek D}. All three of them show a convincing agreement between the model and the empirical data. The  overshooting of the data curve of $Q(t)$ over the model in November and December 2020 seems to  be due to a higher mean time of recorded illness due to an increased number of hospitalized and severe cases. 

\begin{figure}[h]
\includegraphics[scale=0.45]{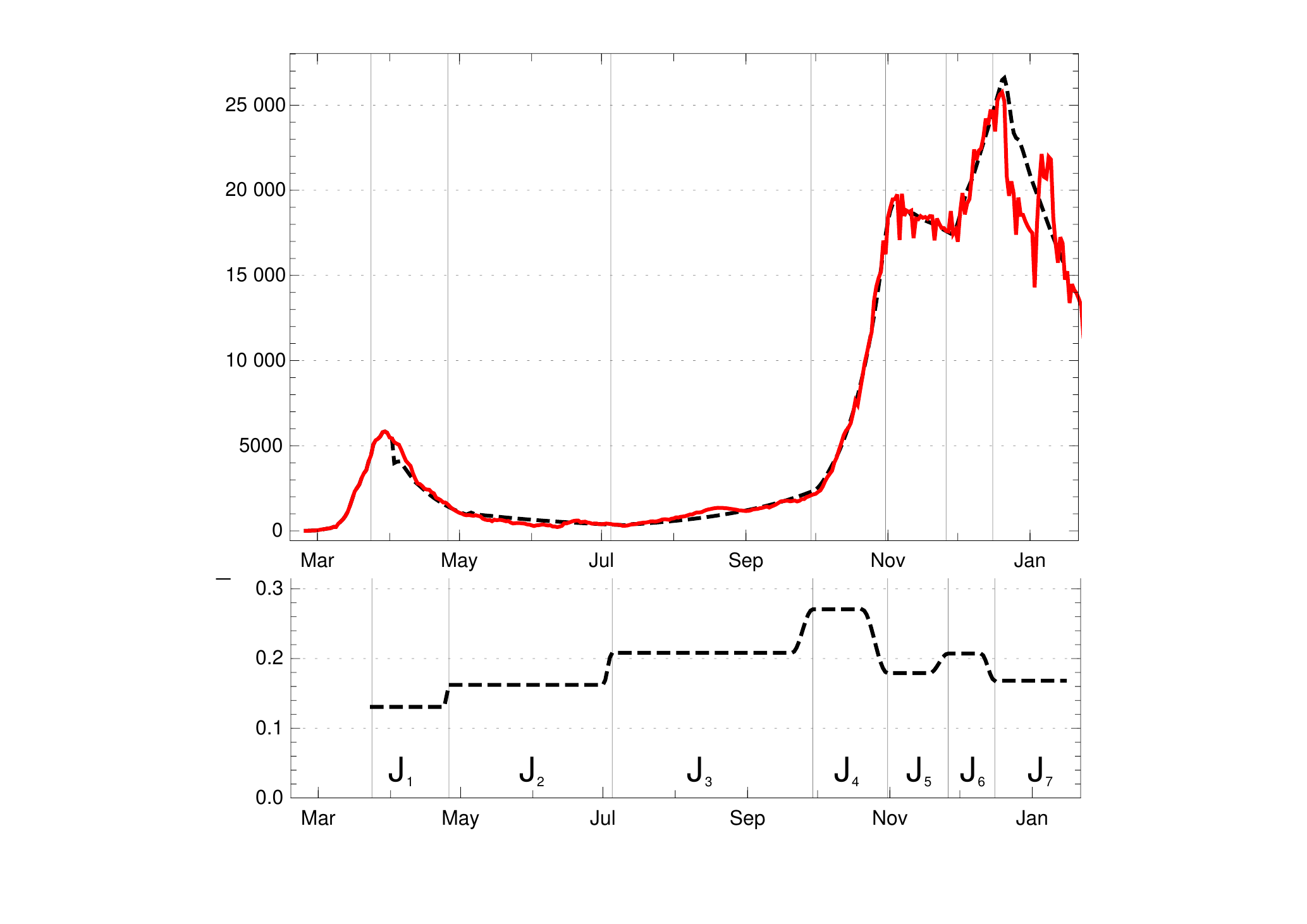}  
\vspace*{-2mm}
\caption{\small Top: Daily new recorded cases $Q_{\text{new}}$  (7-day averages) in Germany; JHU data (solid red) and  model values (black dashed). Bottom:  $\kappa$ used in model, as in fig.~\ref{fig kappa D}   left.
 \label{fig Anew D} }
\end{figure}
\begin{figure}[h]
\begin{center}
\includegraphics[scale=0.6]{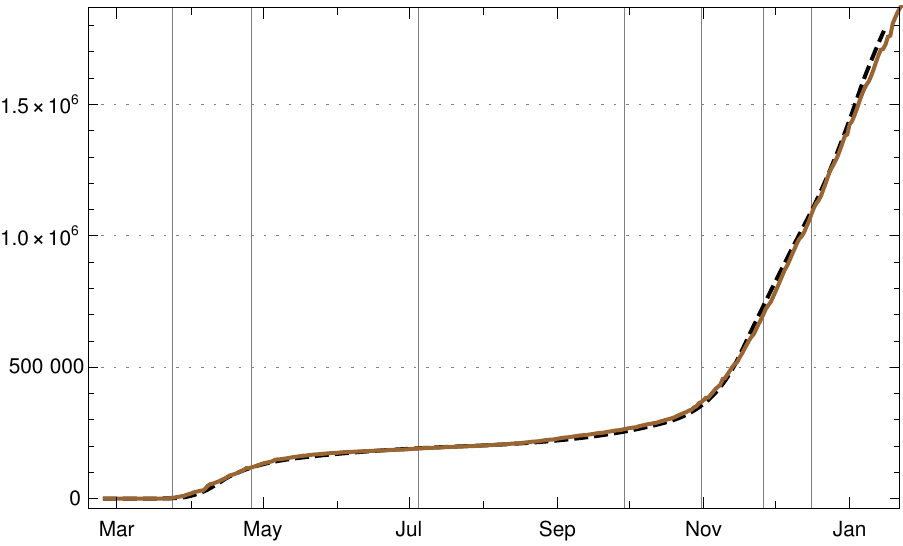} \quad \includegraphics[scale=0.6]{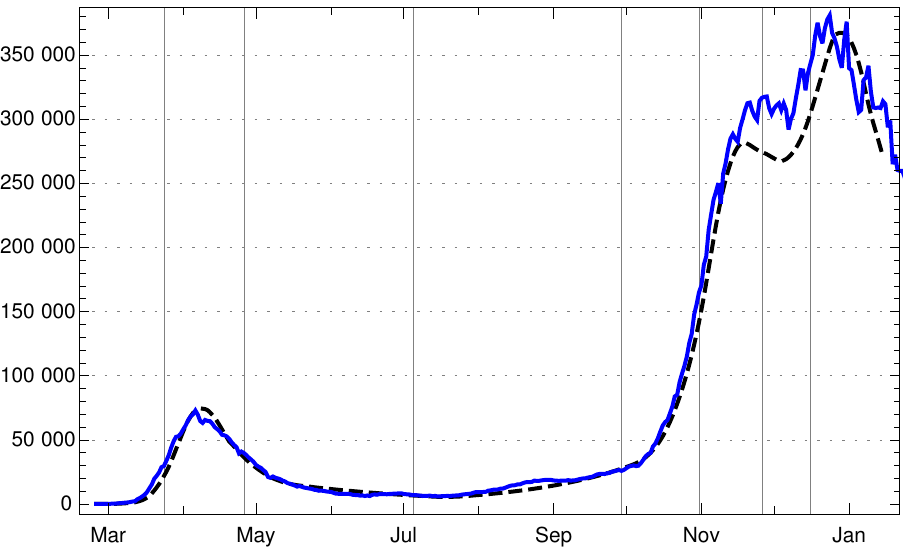}

 \caption{\small Left: Recorded recovered cases $R_c$  in Germany, JHU data (brown) and  model values  (black dashed). Right: Recorded actual cases $Q$ in Germany, JHU data blue, model black dashed. 
 \label{fig Rcrek D} }
 
\end{center}
\end{figure}

\newpage
\section{ A striking application}

Most of the parameters cannot be influenced by politicians, but the contact rates can. Moreover,  there is another one that is highly dependent on social rules, namely the time $p_c$ between the beginning of infectivity and quarantine. One could make attempts to reduce this and check what happens if {\em all other parameters are unchanged, including the contact rate}. 
If we assume that the contact rate is unchanged and we change the value of $p_c$ this will change the model curve. This is clear without any model. But it is one of the strengths of  our model that the input parameters have a direct relation to { the empirical}
 data; in particular $p_c$ shows up as a central parameter. 
 { Obviously }
 a change of $p_c$ results in  a {\em quantitative effect } on the model curves. This effect is drastic. The following graph shows what, according to our model,  would have happened in Germany if from June 2020 on  { the time until quarantine was reduced by 
 only} one day, from $7$ to $6$.

Whereas the actual number of new infections reported per day (7-day average) rose to more than 25000 in December, the model 
{ calculation shows}
that the maximum would have been a bit less than 5000 if the time until quarantine had been reduced in this way.

\begin{figure}[h]
\begin{center}
\hspace{-0.5cm}\includegraphics[scale=0.7]{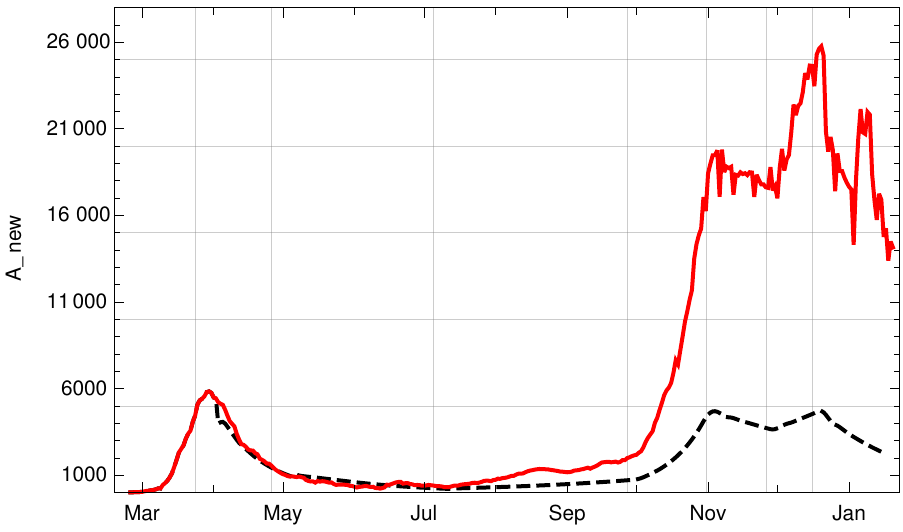}
\caption{Comparison of  7-day averages of recorded daily newly sent to quarantine $Q_{\text{new}}$ in Germany (JHU data, solid red)  with model values (black dashed) assuming a reduction of $p_c$ from 7 to 6; dark sector with $\alpha=0.5$. \label{fig pc 7 auf 6}  }
\end{center}
\end{figure}
\vspace{1em}
\noindent

 \section{Comparison of models \label{section comparison SIR}}
In this section we compare the frequently used standard SIR model with our $\gamma$-K-McK model. Since in the standard SIR model only the three compartments S, I and R occur we reduce the number of compartments in the $\gamma$-K-McK model  by setting $e= 0$, since there is no compartment of exposed in the standard SIR model and $q= 0$, since no quarantine is built in. Then we have to prolong $p_c$ to the length of the interval given by the support of $\gamma$. In addition we remove compartment $P_d$ since  there is no dark sector built into the standard SIR model. To distinguish the functions $S$, $I$ and $R$ of the two models we add the suffix $\mathit{st}$ (like standard) to the functions in the standard SIR model: $S_{st}$, $I_{st}$ and $R_{st}$.  

If one looks at models based on different input functions $\gamma$ it was observed by Fe\ss ler \cite{Fessler:2020} and Diekmann et. al. \cite[sec. 7]{Diekmann_ea:2021},  that there are differences in the long term behaviour if one assumes constant contact rates, even if one is only interested in the number of actually infected people. For short periods the difference is negligible since in all models this is an exponential growth which can be adjusted as long as contact rates don't change. But if non-pharmaceutical interventions are imposed contact rates jump. Thus it is interesting to compare the two models in this situation.

To do this we consider the initial course where $s$ is approximately equal to $1$ and try to adjust the model parameters to achieve a good fit of the model curves. So given a function $\gamma$ with corresponding $\gamma$-K-McK model we can ask whether there are choices of the parameters $\alpha$ and $\beta$ for the standard SIR model so that the model curves essentially agree for some while ($S_{st}'= - \alpha I_{st}S_{st}$ and $R'_{st} = \beta I_{st}$).  In both models the contact rate enters, in the standard SIR model it is proportional to $\alpha$. The role of $\beta$ is often that $\frac 1  {\beta}$ is interpreted as the duration of the infectious period. In our notation this corresponds to $p_c$. But this parameter is not so sharp, for example one can prolong $\gamma$ by a very small constant function so that the support grows but the dynamics determining $S$ is not so much changed. Thus equating $\frac1\beta$ to $p_c$ is somewhat problematic.

For adjusting  the models one might therefore prefer to use the mean generation time $\tau = \frac {\sum_{i} \gamma_i  i }{  \sum_i \gamma_i}$ instead of the duration of the infectious period.\footnote{We thank an anonymous referee for this proposal and Stefan M\"uller for enlightening discussions on this point.}
 If in addition one insists that $N - S$, which approximately is an exponential function at the beginning, agrees for both models then for each $\gamma$-K-McK model one can derive $\alpha$ and $\beta$ such that the mean generation time and the growth rate of $N - S$ or, equivalently, the growth rate of $-S_{st}'$ and $I_{new}$ of the K-McK model agree at the beginning.

With this choice of the  parameters of the SIR model we observe in numerical tests that a change of the contact rate by factor $0.6$  after  50 days shows a very similar course of $-S'_{st}$ and $I_{new}$ of the K-McK mdoel for some time (see fig. \ref{fig model comparison Inew} left).\footnote{Parameters: $\gamma$-function see sec. \ref{section parameters}, $p_d=11,\, \kappa_1=0.228$, $\kappa_2=c \,\kappa_1$ with $c=0.6$, $\tau= 4.47$; initial exponential  growth rate $\eta_1 = \alpha_1-\beta=0.0688$, $\beta=\tau^{-1}=0.224$,  $\alpha_2=c\, \alpha_1$;  $I_0=1000, \, N= 80\cdot 10^6$. \label{fn par 1}}
 Here we chose the parameters $\gamma$  as in the
table in section 3 and $p_c = 11$, the length of the support of $\gamma$. This might be a general fact for reasonable kernels and moderate growth rates as  Stefan M\"uller has pointed out to us.

 \begin{figure}[h]
\begin{center}
\hspace{-0.5cm}\includegraphics[scale=0.7]{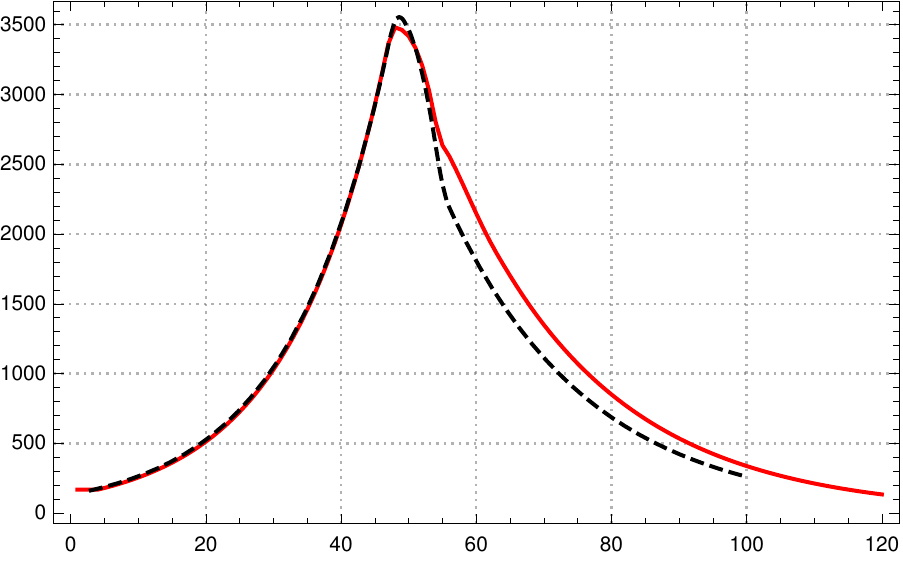}\quad 
\includegraphics[scale=0.7]{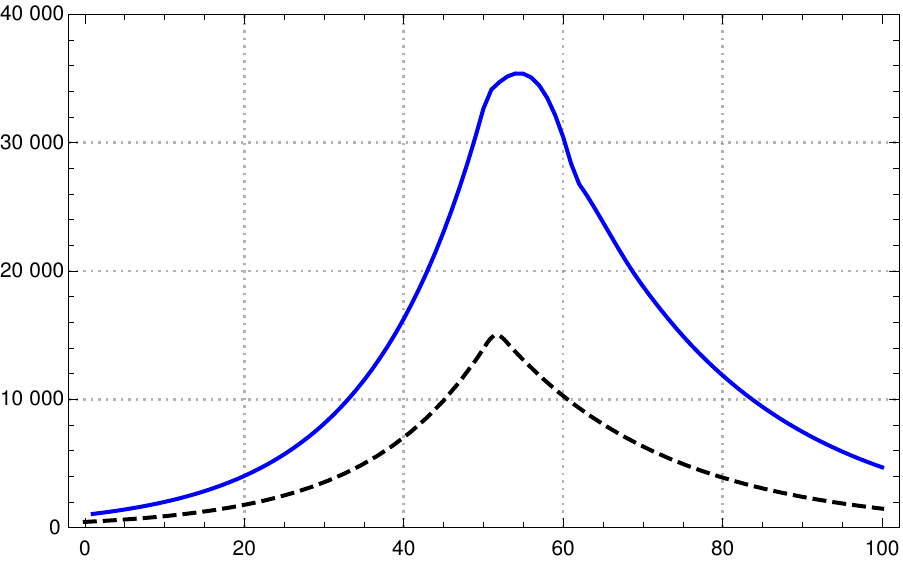}
\caption{Left: Comparison of $-S'_{st}(t)$ for  SIR (black dashed) and $I_{new}(t)$ of the $\gamma$-K-McK model (red) with, approximately, the same initial growth at the beginning. At $t=50$  lowering of contact coefficients, or reproduction rates, by factor $0.6$ ($N=80$ Mill.).  Right: The same for $I(t)$, SIR black dashed, $\gamma$-K-McK blue. 
\label{fig model comparison Inew}  }
\end{center}
\end{figure}

On the other hand, if one wants to control an epidemic one needs to know more than the number of newly infected, which corresponds to   $-S'$, one also wants to know the number $I(k)$ of actually infected (a certain portion of which needs intensive care). So one may ask whether with the above choice for $\alpha$ and $\beta$ besides the curves for $-S'$ also the curves for $I$ are close to each other for both models. Numerical tests show that this is not the case, the curves differ drastically already in the initial course (fig. \ref{fig model comparison Inew} right). 

If one wants to control both $-S'=-(I'+R')$ and $I$,  or  equivalently $I$ and $R$, 
at least in the initial course and then check what happens after changing the contact rates, one asks whether there are values for $\alpha$ and $\beta$ such that during that time both curves agree approximately. This is of course possible, since both curves are approximately exponential functions with same growth rate. Then the question is, what happens under a change of the contact rates. Again we have tested this numerically by changing the contact rate at day 50 by a factor of $0.6$ and found that then {\em both},  $I$ and $R$, diverge drastically (fig. \ref{fig model comparison I R}).\footnote{Parameters of K-McK model and initial exponential growth rate see fn. \ref{fn par 1};  $\beta=0.223$, $ \alpha_1=\beta + \eta_1$,  $\alpha_2=c\, \alpha_1$, $I_0=1000, \, N= 80\cdot 10^6$.   }

The upshot is: If one is only interested in the course of $S$ or $-S'$ numerical calculations indicate that one can choose $\alpha$ and $\beta$ such that as long as $s$ is nearly equal to $1$ both curves are similar even after changes of the contact rates. But the prize one has to pay for this is that then the curves for $I$ diverge even in the initial phase. Alternatively one can chose $\alpha$ and $\beta$ such that in the initial course {\em} $-S'$ and $I$ are nearly the same, but then {\em both} curves diverge after changes of the contact rate. 

So our conclusion is that -- as expected --  the choice of $\gamma$ plays a strong role. Needless to say that we are convinced that a more realistic guess like we made in this paper leads to a better model. This is also supported by our observation indicated at the end of section \ref{section data to model} that the adapted $\gamma$-K-McK model behaves well with regard to all three functions over the range of the epidemic in Germany with several change points. We expect that the same holds if one replaces the medical function $\gamma$ by  the generation time distribution $\zeta$ as in the paper by Sofonea et al. [22]. Also a K-McK model with a Weibull distribution with a shape parameter which suppresses the tail of the distribution sufficiently strong after roughly 12-15 days would not behave too differently.

 \begin{figure}[h]
\begin{center}
\hspace{-0.5cm}\includegraphics[scale=0.7]{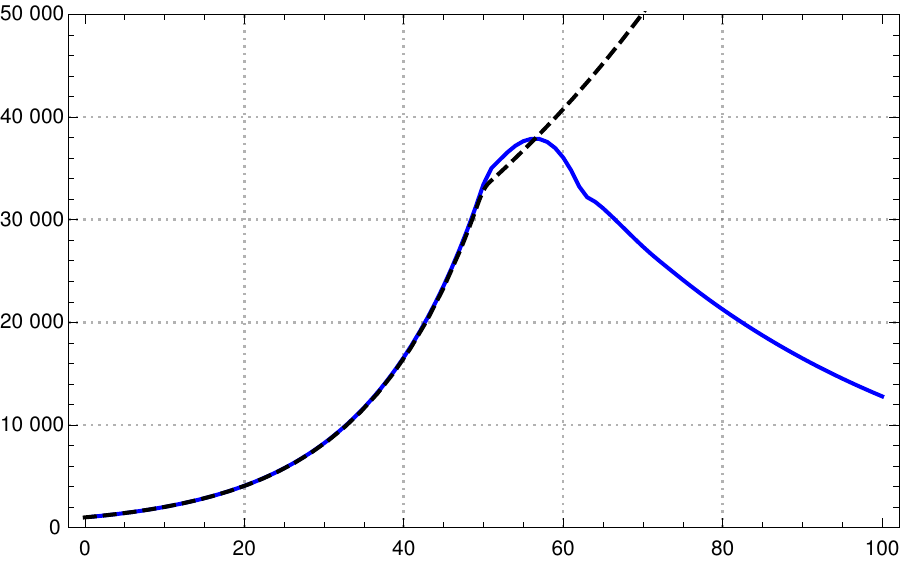}\quad 
\includegraphics[scale=0.7]{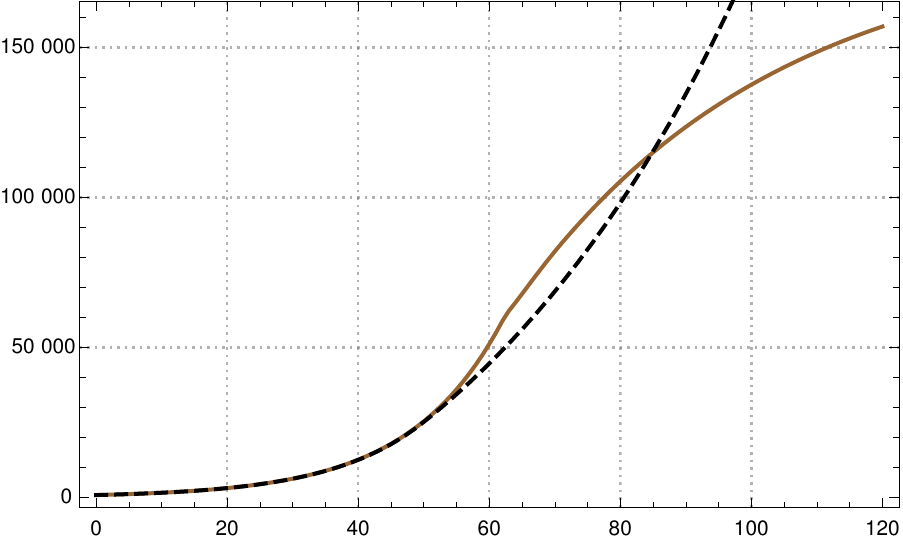}
\caption{Left: Comparison of $I(t)$ for  SIR (black dashed) and $\gamma$-K-McK models (blue) with, approximately, the same initial growth of $I(t)$ {\em and} $R(t)$ (cf. figure on the right). At $t=50$  lowering of contact coefficients, or reproduction rates, by factor $0.6$ ($N=80$ Mill.).   Right: The same for $R(t)$, SIR (black dashed), $\gamma$-K-McK (brown). \label{fig model comparison I R}  }
\end{center}
\end{figure}

We   strongly doubt   that the SIR model can achieve an adequate modelling of all three curves over the whole time of the epidemic, even with complicated manipulations. 
 The more convincing one can relate $\gamma$ (or $\zeta$) to the reality of Covid 19 the more reliable the $\gamma$-K-McK model is.

\section{Vaccination, new mutants, and mass testing \label{section vaccinations}}
In this section we extend our model step by step to include the effects of vaccination, new mutants and mass testing. Please note that all our functions $S(k)$, $E(k)$ etc. have  new definitions in this section and should not be mixed up with the previous formulas. 

Once the past and the expected future rates of daily vaccination are known or estimated,  it is simple to extend the model to take vaccination into account. If  $V_{\text{new}}(k)$ is the number of vaccinated persons on day $k$ and $V(k)$ the total number of immunized by vaccination on day $k$,
 \[ V_{\text{new}}(k)= V(k)-V(k-1)\, ,
 \] 
 equ. (\ref{eq Enew1}) turns into
 \beq E_{\text{new}}(k) = S(k-1)-S(k) - V_{\text{new}}(k) \, .\label{eq Enew3}
 \eeq 
The recursion equation (\ref{eq recursion}) has to be  amended accordingly by the additional summand $ V_{\text{new}}(k)$ on the right hand side (see the Theorem below in the special case  $\zeta = 1$ and $\beta = 0$), and the terms   $\mathbb P_c$  and  $\mathbb P_d$  have to be adapted:
\beqarr
\mathbb P_c (k-1) &:=&  \sum_{j=1} ^{p_c}    \gamma(j) \alpha (k-j)\,  \bigl[S(k-j-e-1) \label{eq PP_c^vm}   \\  \hspace*{7em} && \hspace*{9em}  -S(k-j-e)   - V_{\text{new}}(k-j-e) \bigr]  \nonumber  \\
 \mathbb P_d (k-1) &:=&   \sum_{j=1} ^{_d}  \gamma (j)\, \bigl(1- \alpha (k-j)\bigr) \, \bigl[S(k-j-e-1) \label{eq PP_d^vm}  \\   \hspace*{7em} & &\hspace*{9em}  -S(k-j-e)  - V_{\text{new}}(k-j-e)  \bigr]  \nonumber 
 \eeqarr

Making provision for the appearance of a new mutant of the virus with stronger or weaker infectivity is slightly more involved. 
If we assume that the function of infectivity changes only up to a time dependent scalar factor $\zeta(k)$, with $k$ the epidemic time scale,   $\gamma(j)$ (as above  $j$ denotes  here the day after the onset of infectivity) has to be replaced by
\[
\zeta(k)\, \gamma(j) \, .
\] 
In this way the coefficient of the right hand side of eq. \ref{eq recursion} is  enriched by the factor $\zeta(k)$ (see eq. \ref{eq recursion test}). 

Now we consider (in addition to vaccination and new mutants) reliable unspecific daily mass testing  of a specified subset $M_{\text{t}}$ (with cardinality $N_{\text{t}}$) of the total population $M$ (with cardinality  $N$).
Such testing may be realistic if based on {\em { next} generation sequencing}  as proposed in \citep{Lehrach/Church:2021,Schmidt-Burgk-ea:2020}. We denote the complement of $M_{\text{t}}$   by $M_{\text{n-t}}$ and assume  that the decomposition $M= M_{\text{t}} \cup M_{\text{n-t}}$ is kept fixed over the whole period of  testing. To simplify the considerations we assume that testing takes place  every day and is error free (sensitivity 100\% and specificity 100\%). If in applications the testing starts  later, or takes place on selected days a week only, one has to modify the algorithm appropriately.  We call $\beta := N_{\text{t}}/N$ the {\em test ratio}. Assuming that the tests are reliable implies that there is no dark sector within $M_{\text{t}}$. Moreover, we assume that the test is sensitive from day $l$ of the period of infection on. A positively tested person is assumed to be sent to quarantine on the {\em following} day.

We further assume perfect contact mixing  between $M_{\text{\,t}}$ and  $M_{\text{n-t}}$. If we denote the number of  newly infected at day $k$ amongst people in the tested group $M_{\text{t}}$ by  $E_{\text{t,new}}(k)$ and in the non-tested group by $E_{\text{n-t,new}}(k)$, we have:
\beq E_{\text{t,new}}(k) = \beta \,E_{\text{new}}(k)\, , \qquad \qquad  E_{\text{n-t,new}}(k) = (1- \beta) \,E_{\text{new}}(k) \,  \label{eq E_new^t E_new^n-t}\, ,
\eeq 
where $E_{\text{new}}(k)$ is given by formula (\ref{eq Enew3}).

Now we count the number  $P_{\text{t}}(k)$ of persons from $M_{\text{t}}(k)$  who are effectively propagating the virus under the test regime on day $k$.   Since it takes at most $l$ days until a newly infectious person is either positively or negatively tested we obtain:
\beq P_{\text{t}}(k) = \sum_{j=1}^l E_{\text{t,new}}(k-e-j+1) \eeq  
This formula has to be amended for weekly tests cycles in which tests are not taken daily but only on specified week days.

 We define $Q_{\text{t}}(k)$ as the number of people from $M_{\text{t}}(k)$  who are in quarantine or isolation  on day $k$. Observing that the quarantine starts the day after positive testing we find for the number of people newly sent  to quarantine on day $k$:
 \beqarr Q_{\text{t,new}}(k) &=& E_{{\text{t,new}}}(k-e-l)   \label{eq Q_new test}  \\
  \nonumber
 \eeqarr
 The corresponding number for the people who are not tested is
 \beq Q_{\text{n-t,new}}(k)= \alpha(k - p_c)(1-\beta)\, E_{\text{new}}(k-e-pc) \label{eq Q_n-t,new test}\, ,
 \eeq
 where again $E_{\text{new}}(k)$ is given by formula (\ref{eq Enew3}).
 
Next we define  
 \beqarr \mathbb P_{\text{t}}(k-1) &=& \beta \,  \Big( \sum_{j=1}^l \gamma(j)\, E_{\text{new}}(k-e-j) \Big) \label{mathbb P_t} \\
 \hspace*{3em}&& \hspace{1em}  \, ,\nonumber
 \eeqarr
 where $E_{\text{new}}(k)$ is given by formula (\ref{eq Enew3}).

This number counts only the propagating people who are tested (those in $M_{\text {t}}$); the corresponding number for those who are are not tested (those from $M_{\text{n-t}}$) has to be added. It is essentially given by \ref{eq PP_c^vm} and \ref{eq PP_d^vm}, but the expressions have to be multiplied by the factor $(1-\beta)$, i.e. the percentage of non-tested.  

This allows to formulate our final result including the effects of vaccination, new mutants and mass testing as described above
 \begin{thm}
 If we take vaccination, the rise of new mutants and unspecific mass testing like the above into account, the central recursion of our model becomes
\beqarr
   S(k-1) - S(k)  &=&    V_{\text{new}}(k-1) +  \label{eq recursion test}  \\ \hspace*{1em} && \hspace{-9em}    
   s(k-1)\,  \kappa (k-1) \,  \zeta(k-1)\, \Big( \mathbb P_{\text{t}}(k-1)  +   (1-\beta)[
\mathbb P_c(k-1) + \mathbb P_d(k-1)] \Big) \, ,\nonumber
 \eeqarr
 where $\mathbb P_{\text{t}}$ is given by formula (\ref{mathbb P_t}), $\mathbb P_c$ by (\ref{eq PP_c^vm}), and $\mathbb P_d$ by (\ref{eq PP_d^vm}).

 The number of daily newly detected infected by testing is given by equation (\ref{eq Q_new test}), the ones in the non-tested ensemble  by equation (\ref{eq Q_n-t,new test}). So the total number of new detected infected is
 \beq Q_{\text{new}}(k) = \beta \,E_{\text{new}}(k-e-l) +  \alpha(k - p_c)(1-\beta)\, E_{\text{new}}(k-e-pc) \label{eq Q_new, test}\, ,
 \eeq
 where  $E_{\text{new}}(k)$ is given by formula (\ref{eq Enew3}).
 
Equation (\ref{eq recursion}) is, of course, the special case of equation \ref{eq recursion test} with $\zeta(k) =1$, $\beta = 0$,  and $V_{\text{new}}(k)=0$ (for all $k$).
 \end{thm}

In this case the method for determining the empirical values for the $\kappa(k)$ described in the {\em Observation} above finds the values of the  product $ \kappa (k)\zeta(k)$ rather than $\kappa(k)$ itself. The latter can be isolated by correcting for the factor $\zeta(k)$,  which is to be determined from the data.  
In the following we apply this theorem and study the effect of mutants, of vaccination, and a scenario of mass testing for Germany.

We start with considering the rise of the variant B.1.1.7 in Germany.
It  was first identified in   Germany in late December 2020 and the sequencing data of \cite{RKI:2021mutante} show that the transition ends in early April 2021.   In the German case, we know  that the share of infections carrying the mutant B.1.1.7, determined by genetic sequencing, rose approximately in a linear progression from roughly 6\% in calendar week (CW) 4, the last week of January 2021, to approximately 70\% in CW 10  and 88\% in CW 12 \cite{RKI:2021mutante}.  We may safely assume that about 2 or 3 weeks after CW 10,  i.e. in early  April,  B.1.1.7 will be dominating the infections observed in Germany. 
Given this one has to find  the factor $c$ by which the infectivity increases. This could either be determined by medical studies which we didn’t find or by epidemiological studies. We found two such studies (we would like to thank the referee for these references) in England and France the increase of infectivity of B.1.1.7 has been estimated to be characterized by  a factor $c$ with $1.43 \leq c \leq 1.9$   \citep{Davies_ea:2021} or $1.52 \leq c \leq 1.69$ respectively \citep{Gaymard_ea:2021}.   In agreement with these data we may safely assume that   also in Germany the final scaling factor is constrained by $1.4 \leq c \leq 1.9$.

Using the transition function (\ref{eq g(t)}) introduce above we  model the rise of the factor of infectivity $\zeta$ in dependence of $c$ by the function
\beq \zeta_{c}(t) = 1 + (c-1) g \big(\frac {t-t_a} {d}\big)  \, ,
\eeq
with   $d= t_a-t_b$, where $t_a$ parametrizes the beginning of the spread  of B.1.1.7 in Germany (December 25, 2020),  $t_b$  the date of final dominance  (April 20, 2021), and $c\in [1.4, \, 1.9]$. The case $c=1.5$ is shown in fig. \ref{fig mutante}.

\begin{figure}[h]
\begin{center}
  \includegraphics[scale=0.7]{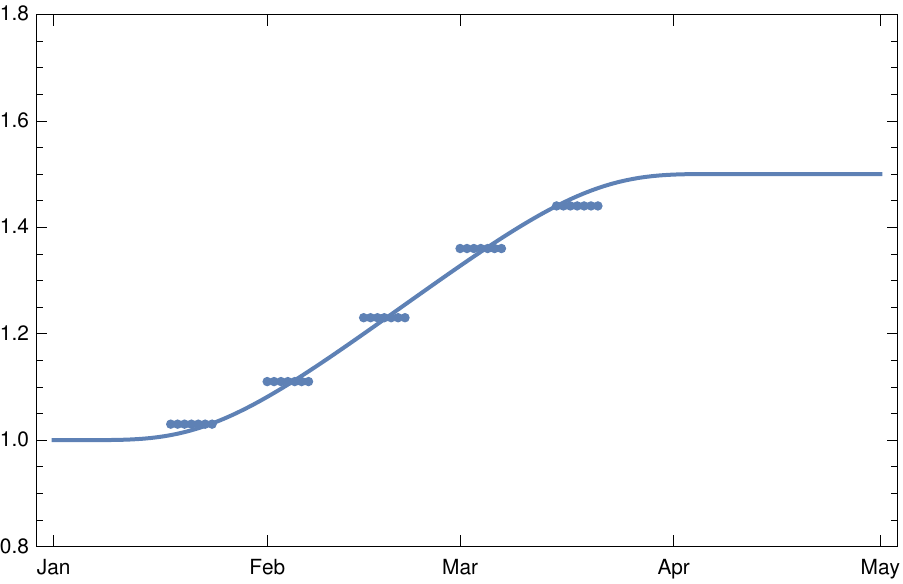}
\caption{\small Left: Scaling function $\zeta_c(k)$  (blue line)  with $c=1.5$  for the infectivity of the mixture of ``old'' virus and new mutants, mainly B.1.1.7 in Germany. The blues dots show the ratio of B.1.1.7  as given in \cite{RKI:2021mutante}, weighted  by $c$ and backward time translated by  7 days. 
 \label{fig mutante} }
\end{center}
\end{figure}

  To constrain $c$  more narrowly  we investigate the  $\kappa(k)$ 
  in early March, determined from empirical data as described in the {\em Observation} of sec. \ref{section parameters}  in dependence of the choice of $c$. The case $c=1$ is shown for comparison  (fig.~\ref{fig mutante2}, top left); it  would correspond to the case of no new mutant. The resulting rise of $\kappa(k)$  would be completely  implausible as no relaxation was announced by the German authorities and the majority of the population accepted the restriction measures issued in December 2020. In early March a  partial opening of shops was permitted and a slightly disorganized discourse started on possible  relaxation of  restrictions. This is sufficient to explain a  moderate rise of $\kappa$ at the beginning of March by 5 - 10 \%, as we find it for $c\approx1.4$ (fig.~\ref{fig mutante2}, top right) and $c\approx 1.6$ (bottom left) . The hypothesis $c\approx 1.9$ appears again implausible, as it would imply that the early March partial opening had no effects on the mean contact rate or  would even be correlated with a small decrease of $\kappa$ (fig.~\ref{fig mutante2}, bottom right).

\begin{figure}[h]
\begin{center}
 \includegraphics[scale=0.6]{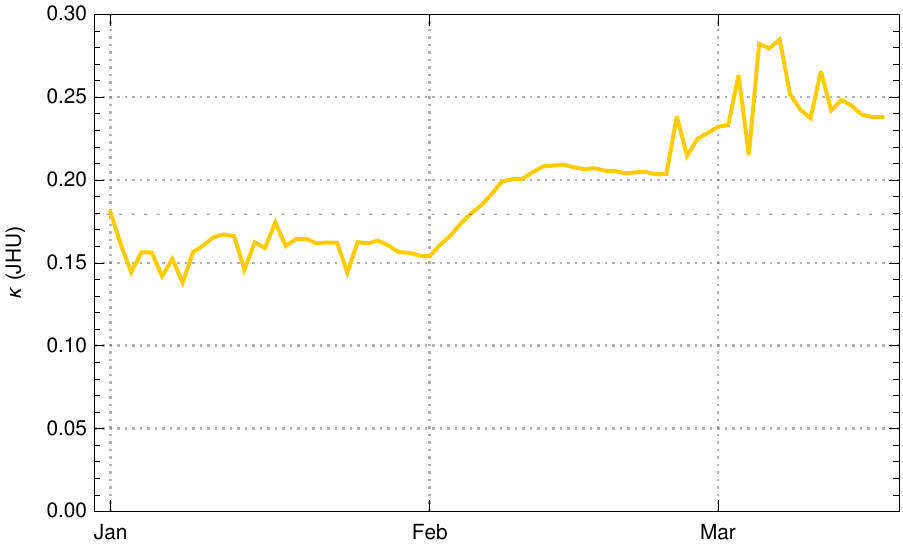}  \quad 
 \includegraphics[scale=0.6]{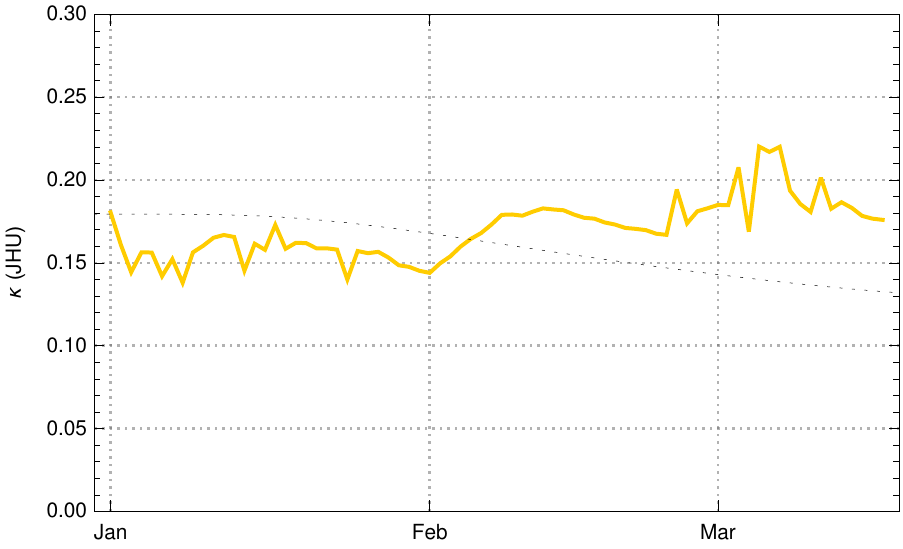} 
 \includegraphics[scale=0.6]{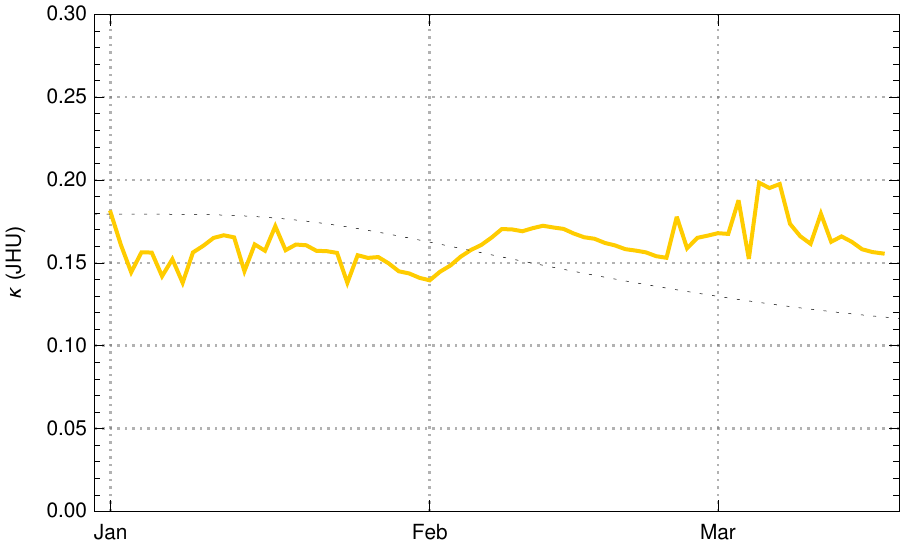}  \quad  \includegraphics[scale=0.6]{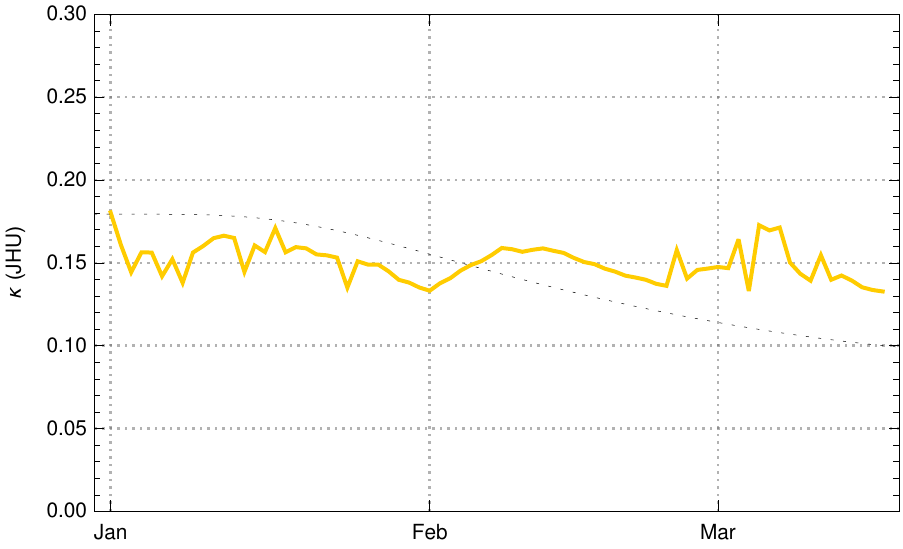} 
\caption{\small Empirical values for  $\kappa(k)$ depending on $\zeta_c(k)$  for Germany (yellow) in  the first three months of 2021 for different values of $c$. Top left: $c=1$  the (counter-factual) hypothesis of no effect due to the new variant. Top  right: $c=1.4$. Bottom left: $c=1.6$. Right $c=1.9$. The dotted downward swung line (top left dotted straight line) shows the critical value  for $\kappa(k)$ (corresponding to the basis reproduction rate = 1).
 \label{fig mutante2} }
\end{center}
\end{figure}

Summing up, we obtain a consistent and plausible choice for the beginning of the transition at December 25, 2020, a duration $d \approx 100$ and an up-scaling factor $1.4 \leq c \leq 1.6$. In the following we work with $c=1.5$.

\begin{figure}[h]
\begin{center}
\includegraphics[scale=0.8]{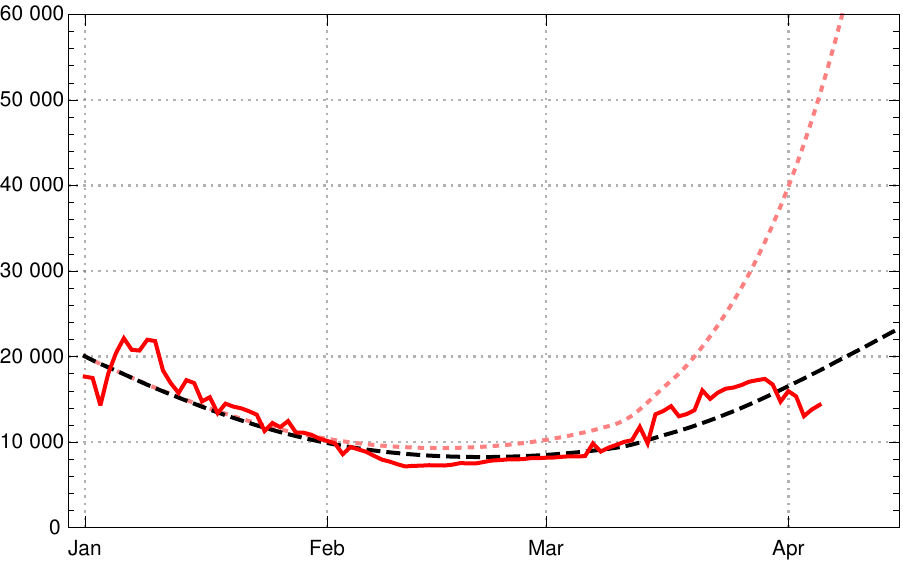}  \quad 
 \caption{\small Daily new infections  for Germany based on data available on April 09, 2021) Red JHU data, black dashed model, pink  dotted model with the same contact rates $\kappa(k)$  and new mutant B 1.1.7 under the (counterfactual) assumption that  no vaccinations had taken place. \label{fig scenario-1 noVAc0} }
\end{center}
\end{figure}

If we  now factor in the available data on vaccinations in Germany,\footnote{\url{https://impfdashboard.de/static/data/germany_vaccinations_timeseries_v2.tsv}} we can analyse the impact of vaccination on the course of the epidemic in Germany. Figure \ref{fig scenario-1 noVAc0} shows the numbers of daily new infections (7-day averages); red the JHU data, black dashed the model reconstruction taking the vaccinations into account, pink dotted  the model values under the  counterfactual assumption that no vaccination had taken place in early 2021.

In fig.~\ref{fig rho Szen1}  one can inspect  the change of the effective reproduction numbers for Germany in early 2021 (data available until end of March) and a { conditional prediction of} the future months April and May under the assumption of no essential changes in the contact rates (orange -- empirical values, black dashed and dotted model with vaccination, pink dotted model without vaccination). Under this assumption  the vaccinations will push the effective reproduction rate  below the critical value 1 at the turn from April to May 2021. If no vaccination were to take place this would happen only in late May -- after a catastrophic increase 
{ in the number of daily new infected people was followed by herd immunity.}

\begin{figure}[h]
\begin{center}
 \includegraphics[scale=0.8]{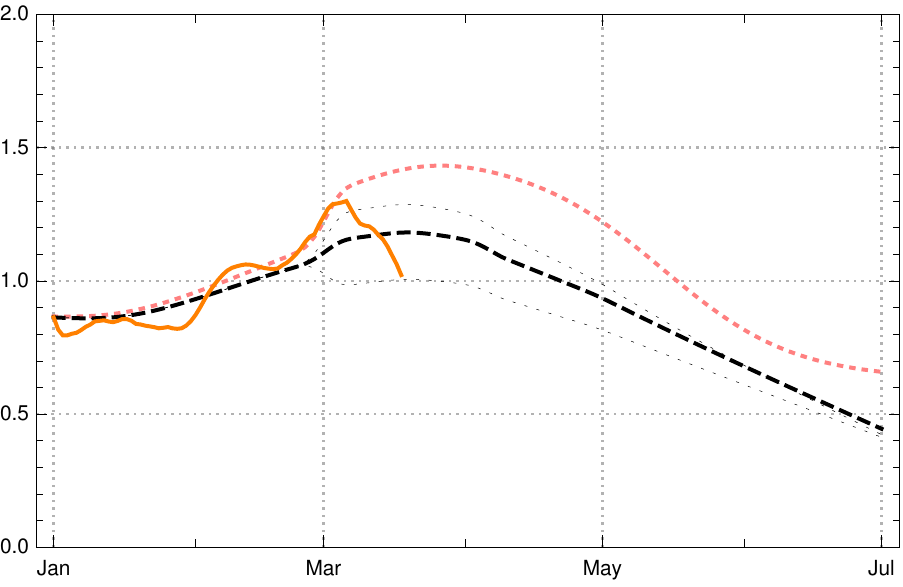}  
  
 \caption{\small Reproduction number $\rho(k)$ for the model scenario above in Germany.   Orange: empirical $\rho(k)$ (using JHU data available until April 09, 2021). Black: model $\rho(k)$; dashed basic scenario, dotted the boundaries of 1 sigma domain of empirical $\kappa(k)$ in the last constancy interval. Pink dotted: model  $\rho(k)$ if no vaccination had happened. \label{fig rho Szen1} }
\end{center}
\end{figure}

We would like to demonstrate the strong effect of  vaccination by showing the catastrophic rise of daily new infections without vaccination,  assuming no essential change of contact behaviour  of the German population (in comparison with March 2021). During April the number of daily new infections would rise above 100.000, see fig. \ref{fig scenario-1 sigma}, dotted pink curve. But because of the ongoing vaccinations our model lets us expect the local maximum of the third wave for daily new infections at the turn from April to May 2021 with values (7-day averages) about 25 000 (black dashed; dotted are the boundaries of the 1 sigma interval for the empirical $\kappa (k)$ in the last constancy interval). Note that the data available on April 09, 2021, are subject to an artificial drop of $Q_{\text{new}}(k)$ induced by  Easter  (April 2--5); this  leads  to additional uncertainties in the data evaluation and conditional predictions. In any case the figures  \ref{fig rho Szen1} and \ref{fig scenario-1 sigma}  demonstrate  the important role of  vaccination in  turning the tide of the epidemic in Germany during the months April and May 2021. 

\begin{figure}[h]
\begin{center}
\includegraphics[scale=0.8]{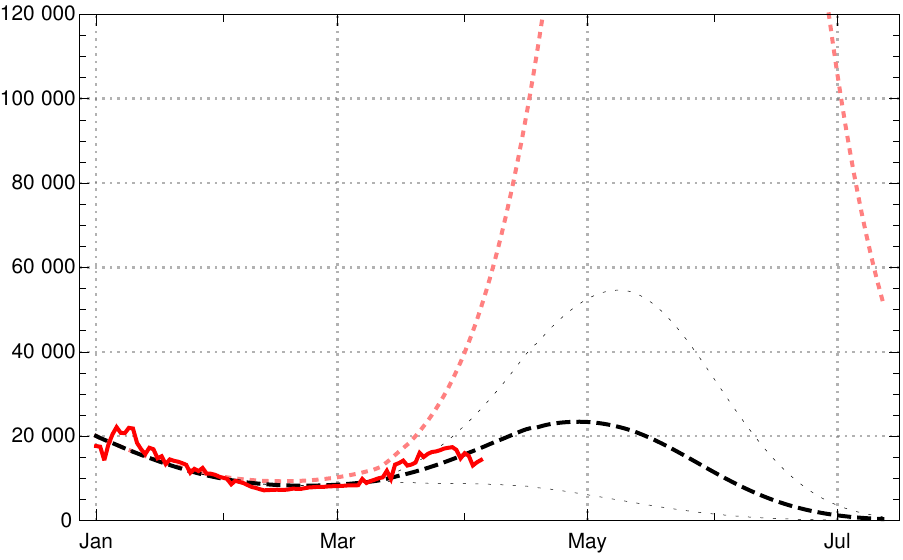}  \quad 
 \caption{\small Future scenario  of daily new infections (7-day averages)  for Germany, based on data available on April 09, 2021, with parameters described in the main text. Red: JHU data. Black dashed: model scenario based on  last available for empirical contact coefficients $\kappa(k)$ ;  black dotted: model scenario  boundaries of 1 sigma domain of empirical $\kappa(k)$. Pink dotted: model scenario under assumption of no vaccination. \label{fig scenario-1 sigma} }
\end{center}
\end{figure}

Finally we apply our  theorem to study the effectiveness of  mass testing techniques, e.g., by  {\em new generation sequencing} proposed in \citep{Lehrach/Church:2021,Schmidt-Burgk-ea:2020}.   The question is  how fast testing on a daily basis of a certain subset  $M_{\text{t}}$ can be expected to suppress an epidemic of the  Covid-19 type.
   We model  here  an ideal mass test (sensitivity and specificity both 100\%) of a fixed subset  $M_{\text{t}}$ containing 60\% of the total population on a daily basis, starting on October 15, 2020. Moreover we even assume (counter-factually) that  the {\em contact regime} of early October is {\em not being changed} by further restrictive measures later in the year, while the effects of  vaccinations and the rise of the new mutant B.1.1.7 since late December 2020 have been included.
   
In other words, we  use the right hand side (rhs) of eq. \ref{eq recursion}  for $t<$ Oct 15  with the contact coefficients of the German model and the rhs of eq. (\ref{eq recursion test}) with constant contact coefficient $\kappa_4$ (see sec. \ref{section data to model}). Using the characteristic function  $f_{\text{test}}(t)= \chi[t, J]$ of the interval $J=[t_a, \infty)$  the recursion is then
\beqarr
   S(k-1) - S(k)  &=&   V_{\text{new}}(k-1)\, +  \\
   \hspace*{8em} & & \hspace{-9em}  s(k-1)\,  \kappa (k-1) \,   \zeta(k-1)\, \Big[ 
    (1-f_{\text{test}}(k-1))  \big(\mathbb P_c(k-1) + \mathbb P_d(k-1) \big)
      \nonumber \\ 
    \hspace*{8em} & & \hspace{-9em} + f_{\text{test}}(k-1)  \Big( \mathbb P_{\text{t}}(k-1)  +   (1-\beta)[\mathbb P_c(k-1) + \mathbb P_d(k-1)] \Big) \Big] \,\nonumber
 \eeqarr
   Figure \ref{fig Lehrach test} shows the  suppression of $Q_{\text{new}}(k)$ which is to be expected  in terms of  our model calculation.

\begin{figure}[h]
\begin{center}
 \includegraphics[scale=0.7]{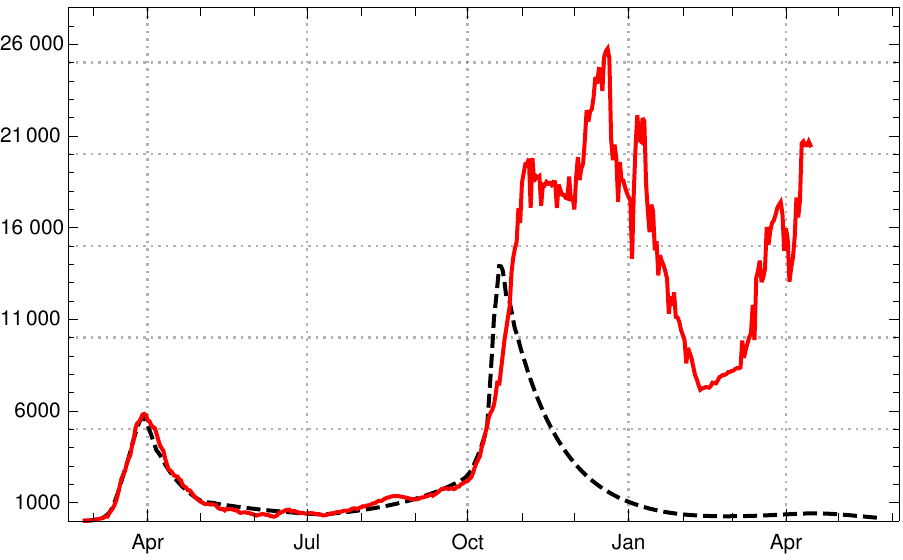}    \quad \includegraphics[scale=0.7]{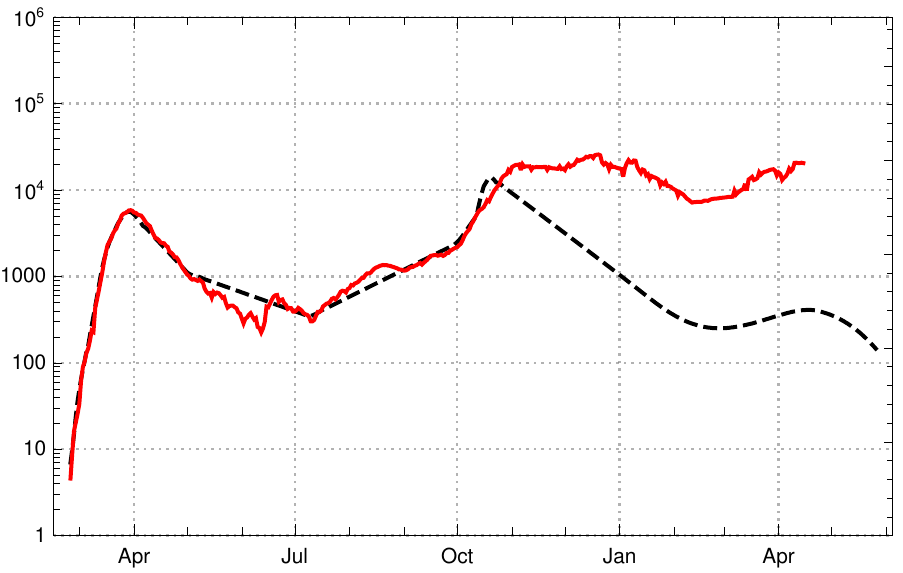}
\caption{\small Comparison of development of registered daily new infections $Q_{\text{new}}(k)$ (7-day averages) of JHU data for Germany (red) with expectation under a test regime with daily testing of 60\% of the population and {\em no} further {\em contact restrictions} than those already  operative in late September 2020 (black dashed). Left: linear scale. Right: logarithmic scale.
 \label{fig Lehrach test}  }
\end{center}
\end{figure}

It demonstrates a high in-principle effectiveness  of such an approach, although in an  idealised scenario. 
Under  a test regime similar to the one above the rise of the new mutant would only lead to a slight increase of registered new infections starting in early March 2021. But it  would be held below roughly 1000  by the increase of vaccinations in April and May (with actual data on vaccination for Germany).  We have checked that for a moderately reduced reliability of the test with  a ratio of 0.9 recognized  infectious persons (sensitivity 90\%)  and daily testing  the picture would not change significantly (the number of false positives  for genetic screening tests is expected to be extremely low; specificity 1 with extremely high precision). 
 
 We would like to remark that the effect of mass testing has a similar effect as the shortening of the time between the offset of symptoms and sending people to quarantine or isolation as discussed in our striking application. Mass testing is a particular effective way to shorten this period for a certain part of the population.

\section{Comparison of reproduction numbers \label{section comparison rep}}

The  Robert Koch-Institut (RKI), Berlin,    publishes the daily data for an epidemic in Germany and determines the daily reproduction numbers from these data. 
The RKI calculation uses a method  described in \cite{anderHeiden_ea:2020} for a stochastic estimation of the numbers of newly infected, called $E(t)$, from the  raw data of newly reported cases,.
The calculation of the reproduction numbers works with these $E(t)$ and assumes  constant generation time and serial intervals of equal lengths $\tau_g=\tau_s=4$
\cite{RKI:2020Erlaeuterung}.
Two versions of reproduction numbers are being used, a day-sharp and therefore ``sensitive'' one 
$\rho_{rki,\, 1} (t) = \frac{E(t)}{E(t-4)}$,
  and a  weekly averaged one,
\[ \rho_{rki,\, 7} (t) = \frac{\sum_{j=0}^6 E(t-j)}{\sum_{j=0}^6 E(t-4-j)} \, , 
\]
which we refer to in the following simply as $\rho_{\text{RKI}}(t)$. 

The paper remarks that the  RKI  reproduction numbers (``$R$-values'') $\rho_{\text{RKI}}(u)$  indexed by the date $u$ of calculation  refer to a period of infection which, after taking the incubation period $\iota$ between 4 and 6 days into account, lies between $u-16, \ldots, u-8$ (with  central day $u-12$ in the interval). The  reproduction numbers calculated  in the   adapted K-McK approach are  close to the RKI reproduction numbers.  The main differences lie in the usage of different raw data bases (RKI versus JHU) and the adjustment of the raw data (stochastic redistribution $E(t)$ versus sliding 7-day averages $\hat{Q}_{new,7}$). After a reasonable time shift the agreement between the reproduction numbers of the adapted $\gamma$-K-McK approach and  the  $\rho_{\text{RKI}}(k)$ are close. The stochastic smoothing of the RKI data seems to lead to smaller amplitudes of the fluctuations which otherwise are in striking coincidence (fig.~\ref{fig rho-RKI rho7}).

\begin{figure}[h]
\begin{center}
 \includegraphics[scale=0.8]{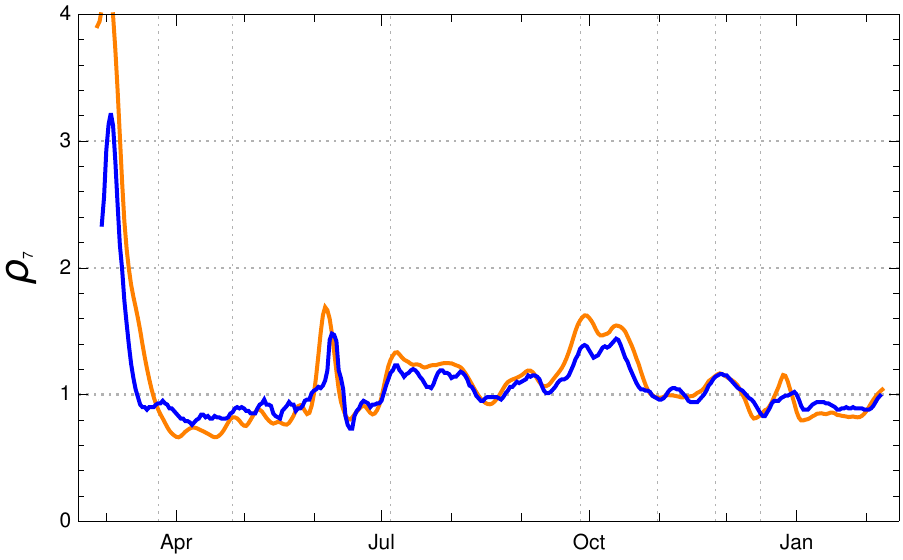} 
\caption{\small 
Empirical reproduction numbers $\rho(k)$ of the adapted $\gamma$-K-McK model for Germany (orange) and reproduction numbers $\rho_{\text{RKI}}(k -9)$  of the RKI (blue); both of them 7-day averages.
 \label{fig rho-RKI rho7} }
\end{center}
\end{figure}

\section{Discussion} 
We formulate four  points with regard to the adapted K-McK-model, 
  which at present seem  the most important ones to us. 
  
 (1) A  model has to be useful for analysing an epidemic on the basis of  the available data. The most important datum for this purpose is the number of newly infected people each day; but one is interested in other numbers also, such as the number of actually counted as infected  or those in hospital, or how many people are actually infectious or how many people are already immune. In particular one is interested in the daily reproduction number (see formula (\ref{eq rho}). All these numbers are given by our model.  Knowing these numbers is  important information for  politicians as well as for citizens. For politicians such numbers provide a basis for thinking about regulations and for citizens this information is a helpful guide for how to behave.

\begin{itemize}
\item[ ] {\em   The present  model is a useful tool for determining key figures of an epidemic such as the reproduction number and the number of people in quarantine or hospital.} 
\end{itemize} 

(2) A reliable model can do more. The optimum would be that it  allows one to predict the future.  But in the strict sense this is impossible, since important parameters such as  the contact rate changes depending on the behaviour of the population. If one knows the development of the epidemic in the past (in terms of a model)  and assumes  that the contact rate is not changing drastically,  it is possible  to extrapolate the development for some time. We indicate how this may be done by looking at the situation in Germany at the beginning of April 2020. In a series of preliminary steps a lockdown was imposed on March 22. Only a few days later it pushed the reproduction number below 1, and about a week later  down to about $\rho = 0.7$ (see fig.~\ref{fig kappa D}). It fluctuated around that number   for roughly a month. This  was observable in the data about a fortnight  later, so that around April 10 the number was stable already for about a week. If at that moment one were to assume that the behaviour of the population is not changing very much, one could use the model to predict the future for the next weeks quite successfully until the contact rate was changing substantially. Figure \ref{fig Anew D}   shows how well this worked until the end of April when the restrictions were partially relaxed and the reproduction number rose.  In other words:

\begin{itemize}
\item[ ]
{\em The adapted K-McK-model is a reliable tool for extrapolating the development in times where the reproduction number is stable and there are reasons to assume that this stays so for a while.}
\end{itemize}

(3) At the beginning of the epidemic there was no evidence of  what the effect of interventions would be. But over the time there was a chance of using the model to observe  the effects and, if possible, to draw realistic conclusions.  A quantitative estimate of the effectiveness of non-pharmaceutical interventions in terms of  the change of  contact rates (or reproduction numbers) determined in the model framework is a delicate point.  But keeping this in mind,  a realistic model such as the  adapted K-McK-model  may be used heuristically for  deriving  useful information about how certain interventions will influence the development by comparison with other ones in the past which took place under similar conditions. 
 Like any realistic model,  the present one may be used as a tool for learning how certain interventions under specified conditions (such as climate) influence the development.
 This was also tried in \cite{Dehning-ea:2020} for the three interventions in Germany in March 2020,  claiming a direct relation between each of these and a new level of the reproduction number. In the light of the data evaluation represented in figure (\ref{fig kappa D}) we are not able to confirm this result. In the  decline of the contact rates, respectively reproduction numbers, in Germany during March 2020 no clear 
  intermediate steps can be discerned which could be read as the signature 
     of the first two interventions. To assume the existence of constancy intervals  between two non-pharmaceutical interventions per se  seems  doubtful to us. In this phase we see only a cumulative effect of all three interventions.

On the other hand, if the data evaluation shows a relative stability of the contact rate $\kappa(k)$, a   conditional prediction for the coming development is possible, sometimes even for several weeks. ``Conditional'' means here under the  assumption of  no essential change of the  behaviour of  people, relative stability of climatic conditions and no mutation of the virus. In section \ref{section data to model} time intervals in which such conditional predictions could be made have been discussed. For example, one could recognize a stable trend of increasing numbers of daily newly infected people for Germany  already in late July and early August 2020 and could have foreseen the approach of the second wave long before  autumn (see fig.~\ref{fig Anew D}, bottom,  and table in Section~\ref{section data to model}).

{
In this context we would like to point to two papers which are of interest in this context. The authors of \citep{Flaxman_ea:2020} study the effects of non-pharmaceutical interventions on COVID-19 in 11 European countries. It would be very interesting to apply our model to the data  and compare the results with theirs. The other paper studies how epidemic control can be obtained by contact tracing \citep{Ferretti_ea:2020}. If such data are available one could improve the application of our model by improved estimates of the contact rates. 
}

  As    conditional predictions with the present model turn out to be convincing and reliable, we also consider future scenarios  for  the overall  development of the epidemic for up to a few months as informative. For an example of a scenario that can be checked against the real course of the epidemic see the end of section \ref{section vaccinations}. 
Such scenarios may be helpful  for exploring  alternative actions (e.g.~non-pharmaceutical interventions) and for identifying necessary  steps to avoid the breakdown of the health system.
 
 \begin{itemize}
 \item[ ] {\em 
 The study of future scenarios with the  $\gamma$-K-McK model  may be useful for persons or institutions   looking for orientation with regard to alternative  actions and/or external changes of condition (rise of new mutants, climate change).
 }
 \end{itemize}
 
(4) As we have seen,  there is a parameter which can, in principle, be influenced by socio-political decisions  comparatively easily, the time between infection and the day people go to quarantine. We therefore emphasize again:

\begin{itemize}
\item[ ]
 {\em The model shows that reducing the time until entering quarantine by one day leads to a drastic improvement.  }
\end{itemize}

 Whether one can reduce it depends on various factors, in particular the infrastructure and effectiveness of the health system. Discussions with experts about the German health system have convinced us that there are good chances for lowering the time until quarantine by at least one day. Of course, this needs a great effort: One has to enable the health system to carry it out and one has to convince the population to follow the corresponding rules. But the latter should have a good chance of success since this is a restriction which  hits only a small number of the population: those who show first symptoms or are identified as Covid-positive in an unspecified test.

(5) A comparison between different models is difficult. If models are based on clear principles the first thing one could do is to compare the principles and discuss their strengths and weaknesses. If one wants to compare the standard SIR model with our adapted model this is simplified by the fact that they are both based on the same principle, the Kermack-McKendrick idea. So the first think is to compare the input functions. But it still could be that although they are very different both models are approximately equivalent. In section \ref{section comparison SIR} we have shown that this is not the case. If one takes both the compartments $I$ and $R$ into account ($S$ is determined by $I$ and $R$), the difference is drastic. We conclude that a model based on a more realistic $\gamma$ function, like the one we have chosen, seems to be more useful.  

 Saying this, one should not forget that no model is a mirror of  reality. It  can even happen that a model based on less realistic principles may work better, but so far we don't see a sign of this in the comparison of the adapted model with the standard SIR model. So we conclude:

\begin{itemize}
\item[  ] {\em Our comparison of the adapted model with the standard SIR model indicates that the former one is based on more realistic assumptions. Since the differences in the long run or when the contact rates jump are large, the adapted model should be the first choice.}

\end{itemize}

\vspace{2em}
\section*{Acknowledgments}

We thank Odo Diekmann for discussing our thoughts. His help was invaluable to us, non-experts in the field, for coming to a proper understanding of compartment models. Similarly we thank Stefan M\"uller for extensive discussions in particular concerning the comparison of models. Section 6 is greatly influenced by thoughts he and one of the referees shared with us. Moreover, we thank Martin Bootsma, Robert Fe\ss{}ler, Jan Mohring, Robert Schaback,  Hans Lehrach { for valuable discussions, the  referees for very useful hints, 
Jeremy Gray for tidying up our English  of the final version,
 }
   and  Don Zagier for his intensive support in the  preparation of the manuscript. \\ Calculations and graphics have been made with Mathematica 12.

\small

\end{document}